\documentclass[aps,showpacs,amsmath,amssymb,twocolumn,pra,superscriptaddress,notitlepage]{revtex4-2}

\usepackage{qcircuit}
\usepackage{amsmath,bm}
\usepackage[dvips]{graphicx}
\usepackage{amsmath,amssymb,amsthm,mathrsfs,amsfonts,dsfont}
\usepackage{subfigure, epsfig}
\usepackage{braket}
\usepackage{bm}
\usepackage{enumerate}
\usepackage{color}
\usepackage{graphicx}
\usepackage{algorithm}
\usepackage{algorithmic}
\usepackage{braket}
\usepackage{comment}
\usepackage{appendix}
\usepackage{here}
\usepackage{tabularx}
\usepackage{physics}
\usepackage{mathtools}
\newcommand{\red}[1]{\textcolor{black}{#1}}
\newcommand{\redd}[1]{\textcolor{black}{#1}}
\newcommand{\reddd}[1]{\textcolor{black}{#1}}

\begin{document}

% -------------------- TITLE --------------------

\title{Virtual quantum error detection}

% ------------ AUTHORS AND AFFILIATIONS ----------

\author{Kento Tsubouchi}
\email{tsubouchi@noneq.t.u-tokyo.ac.jp}
\affiliation{Department of Applied Physics, University of Tokyo,
7-3-1 Hongo, Bunkyo-ku, Tokyo 113-8656, Japan}

\author{Yasunari Suzuki}
%\email{}
\affiliation{NTT Computer and Data Science Laboratories, NTT Corporation, Musashino, 180-8585, Tokyo, Japan}

\author{Yuuki Tokunaga}
%\email{}
\affiliation{NTT Computer and Data Science Laboratories, NTT Corporation, Musashino, 180-8585, Tokyo, Japan}

\author{Nobuyuki Yoshioka}
%\email{}
\affiliation{Department of Applied Physics, University of Tokyo,
7-3-1 Hongo, Bunkyo-ku, Tokyo 113-8656, Japan}
\affiliation{Theoretical Quantum Physics Laboratory, RIKEN Cluster for Pioneering Research (CPR), Wako-shi, Saitama 351-0198, Japan}
\affiliation{JST, PRESTO, 4-1-8 Honcho, Kawaguchi, Saitama, 332-0012, Japan}

\author{Suguru Endo}
\email{suguru.endou.uc@hco.ntt.co.jp}
\affiliation{NTT Computer and Data Science Laboratories, NTT Corporation, Musashino, 180-8585, Tokyo, Japan}

% -------------------- ABSTRACT --------------------

\begin{abstract}
%Quantum error correction and quantum error detection necessitate syndrome measurements to detect errors. Syndrome measurements need to be performed for each stabilizer generator with single-shot measurements, which can be a significant overhead, considering the fact that the readout fidelity is generally lower than gate fidelity in the current quantum hardware. Here, by generalizing a quantum error mitigation method called symmetry expansion, we propose a protocol that we call virtual quantum error  detection (VQED). This method {\it virtually} allows for evaluating computation results corresponding to post-selected quantum states obtained through quantum error detection during circuit execution without syndrome measurements. Furthermore, while the conventional quantum error detection needs the implementation of Hadamard test circuits for each stabilizer generator, our VQED protocol can be performed with a constant depth shallow quantum circuit with an ancilla qubit, irrespective of the number of stabilizer generators. 
%In addition, \red{the obtained computation results are robust against noise in the operation of VQED, and our method}
%VQED
%is fully compatible with other error mitigation schemes for further improvement of computation accuracy, thus leading to high-fidelity quantum computing. 
Quantum error correction and quantum error detection necessitate syndrome measurements to detect errors.
\red{Performing syndrome measurements} for each stabilizer generator can be a significant overhead, considering the fact that the readout fidelity \red{in the current quantum hardware} is generally lower than gate fidelity.
Here, by generalizing a quantum error mitigation method \red{known as} symmetry expansion, we propose a protocol \red{called} virtual quantum error detection (VQED).
This method {\it virtually} allows for evaluating computation results corresponding to post-selected quantum states obtained through quantum error detection during circuit execution, \red{without implementing syndrome measurements}. 
\red{Unlike conventional quantum error detection, which requires} the implementation of Hadamard test circuits for each stabilizer generator, our VQED protocol can be performed with a constant depth shallow quantum circuit with an ancilla qubit, irrespective of the number of stabilizer generators. 
\red{Furthermore, \reddd{for some simple error models,} the computation results obtained using VQED are robust against the noise that occurred during the operations of VQED, and our method} is fully compatible with other error mitigation schemes\red{, enabling further improvements in computation accuracy and facilitating} high-fidelity quantum computing.

%Quantum error correction and quantum error detection require syndrome measurements to detect errors.
%Quantum error correction and quantum error detection necessitate syndrome measurements to detect errors.
%Performing syndrome measurements for each stabilizer generator can be a significant overhead, considering the fact that the readout fidelity in current quantum hardware is generally lower than the gate fidelity. To address this issue, we propose a protocol called virtual quantum error detection (VQED) by generalizing a quantum error mitigation method known as symmetry expansion.
%The VQED protocol {\it virtually} allows for evaluating computation results corresponding to post-selected quantum states obtained through quantum error detection during circuit execution, without the need for syndrome measurements. Unlike conventional quantum error detection, which requires the implementation of Hadamard test circuits for each stabilizer generator, our VQED protocol can be performed using a shallow quantum circuit with a constant depth and an ancilla qubit, regardless of the number of stabilizer generators.
%Furthermore, the computation results obtained using VQED are robust against the noise that occurred during the operation of VQED, and our method is fully compatible with other error mitigation schemes, enabling further improvements in computation accuracy and facilitating high-fidelity quantum computing.
\end{abstract}

\maketitle
%\onecolumngrid
\section{introduction}
\label{section: introduction}
The last decade has seen the remarkable development of the noisy-intermediate quantum computing paradigm from both theoretical and experimental sides~\cite{preskill2018quantum,mcardle2020quantum,cerezo2021variational,tilly2022variational,bharti2021noisy,arute2019quantum,kandala2017hardware,madsen2022quantum}. Nevertheless, the effect of noise lies as a crucial problem in realizing practical quantum computing. Quantum error correction (QEC) and quantum error detection (QED), which reduce computation errors through the encoding of logical qubits with many physical qubits, have been investigated for enhancing computation accuracy for a long time since the early days of quantum information science~\cite{devitt2013quantum,lidar2013quantum,grassl1997codes,steane1996error,shor1995scheme,laflamme1996perfect}. Syndrome measurements \red{are performed in QEC and QED to detect physical errors by using ancilla qubits}; QEC actively corrects physical errors based on the error information obtained in the decoding process while QED discards the noisy quantum states once an error is detected. 

While the utility of QEC and QED have been shown theoretically in numerous previous works, they require high-fidelity syndrome measurements of stabilizer generators. Furthermore, the number of required syndrome measurements increases \red{with} the number of stabilizer generators \red{in} the QEC/QED code. Considering the current situation of superconducting hardware, in which the measurement fidelity is lower than gate errors~\cite{hicks2022active,gunther2021improving,arute2019quantum}, the necessity of single-shot measurements~\footnote{\red{In this work, we use the term ``single-shot measurement'' to represent the measurements that are performed only once, rather than repeating the measurement many times in order to obtain the expectation value of some observable. Note that the meaning is different from the term ``single-shot error correction''~\cite{bombin2015single}.}} for syndrome measurements can be a significant overhead in QEC/QED. 

For the ease of error reduction in near-term quantum hardware, a class of error reduction techniques referred to as quantum error mitigation (QEM) has been recently studied~\cite{temme2017error,li2017efficient,endo2018practical,endo2021hybrid,cai2022quantum}. \textcolor{black}{In many QEM methods, the noiseless expectation values of observables are estimated via post-processing of measurement results.} \red{This indicates that we cannot physically obtain quantum states with reduced noise; nevertheless, QEM allows for {\it virtually } simulating the expectation values of observables for such states.}
Symmetry expansion (SE) is one of the QEM methods that use  symmetries inherent to the system \red{to mitigate errors}~\cite{bonet2018low,mcclean2020decoding,cai2021quantum,endo2022quantum}. The noisy quantum state is virtually projected onto the symmetric subspace \red{through} random sampling of the symmetry operators, additional measurements, and classical postprocessing of measurement outcomes.  As we will discuss later, SE allows for the calculation of the expectation value of an observable corresponding to the post-selected quantum states through QED without implementing syndrome measurements, and hence is suitable for near-term hardware.
So far, SE is theoretically formulated for error mitigation for noisy states immediately before measurement~\cite{mcclean2020decoding,cai2021quantum} and state preparation for rotation symmetric bosonic codes~\cite{endo2022quantum}.
\red{Thus, SE in its current form cannot effectively suppress the accumulation of noise during computation,}
whereas the conventional QED can be more flexibly used during the circuit execution. 

In this work, we significantly expand the framework of SE so that it can be leveraged during the execution of quantum algorithms. Because our method enables us to obtain the expectation values corresponding to the post-selected state via QED, we call it {\it virtual quantum error detection} (VQED). 
\red{While the conventional SE can only detect errors immediately before the measurement of expectation values, VQED can detect errors even while the execution of the quantum circuit, enabling us to mitigate the accumulation of errors during the computation.}
Although VQED inherits the disadvantages of SE, i.e., we can only obtain error-mitigated expectation values, not the quantum state itself, and the required sampling complexity is quadratically worse for the success probability of QED, the significant advantages of VQED compared with the QED are as follows: 1. the depth for QEM is constant regardless of the number of stabilizer generators of the code; 2. we only need to measure an expectation value of an observable without the need for single-shot syndrome measurements;
\red{3. the obtained expectation values are robust against the noise that occurred during the operations of VQED \reddd{for some simple error models};}
\red{4}. our method is fully compatible with other QEM methods, e.g., readout error mitigation~\cite{maciejewski2020mitigation, bravyi2021mitigating} for the ancilla qubit used in our protocol.
We numerically verify the behavior of the fidelity improvement with our VQED protocol over the conventional SE and the unencoded \red{physical qubits}.
\red{We also evaluate the required sampling costs and verify that the sampling cost for VQED does not significantly increase compared to SE.}
%and evaluate the required sampling costs.We remark that 
\red{Furthermore}, our method can offer virtual implementation of stabilizer-like QEM methods using spin and particle number preservation in the computation~\cite{mcardle2019error,bonet2018low} in even more hardware-friendly manner.

In addition, we discuss the virtual implementation of quantum error correction, which results in the computation outcome corresponding to the error-corrected quantum states. While the conventional QEC does not induce additional sampling overheads, we find that our virtual QEC generally incurs a larger sampling overhead than the virtual QED method; therefore, we conclude VQED is preferred in typical quantum computation scenarios.

\section{Preliminaries}

\subsection{\textcolor{black}{Quantum} error detection and \textcolor{black}{quantum }error correction for stabilizer codes}
\label{sec_1}
We first review stabilizer codes and ways to detect and correct their errors~\cite{nielsen2002quantum,gottesman1997stabilizer}.
QED and QEC are performed by encoding quantum information into enlarged Hilbert space at the expense of multiple quantum systems. Due to its redundancy, we can detect and correct their errors during the computation.

Here, we review the stabilizer formalism, which is the most standard method to construct quantum error-correcting codes.
Consider an $n$-qubit Pauli group as
\begin{eqnarray}
    \mathcal{G}_n = \qty{\pm1, \pm i}\times \qty{I, X, Y, Z}^{\otimes n}
\end{eqnarray}
where $I$ is the identity operator for single qubit system and $X =\begin{pmatrix}0&1\\1&0\end{pmatrix}$, $Y =\begin{pmatrix}0&-i\\i&0\end{pmatrix}$, and $Z =\begin{pmatrix}1&0\\0&-1\end{pmatrix}$ are Pauli operators.
To encode $k$ logical qubits into $n$ physical qubits, we define a stabilizer group $\mathcal{S} = \qty{S_1, \cdots, S_{2^{n-k}}} \subset \mathcal{G}_n$ as a commutative subgroup of the Pauli group $\mathcal{G}_n$ with $-I^{\otimes n}\notin \mathcal{S}$. 
We denote a generator set of the stabilizer group $\mathcal{S}$ as $\mathcal{G} = \qty{G_1, \cdots, G_{n-k}}$.
Then, we can define the logical space of the stabilizer code $\mathcal{C}$ as an eigenspace with $+1$ eigenvalues for all the operators in the stabilizer group, i.e., $\mathcal{C} = \qty{\ket{\psi}|\forall S_i \in \mathcal{S}, S_i\ket{\psi} = \ket{\psi}}$.
In the $2^{k}$-dimensional Hilbert space, we can introduce a logical basis as $\qty{\ket{0}_L, \ket{1}_L}^{\otimes k}$ and logical Pauli operators as $\qty{I_L, X_L, Y_L, Z_L}^{\otimes k}$.
The code distance $d$ is the minimum number of physical qubits on which an arbitrary logical operator of the code non-trivially operates. We denote such stabilizer codes as \red{$[[n,k,d]]$} stabilizer codes.

\begin{figure}[t]
    \begin{center}
        \includegraphics[width=0.9\linewidth]{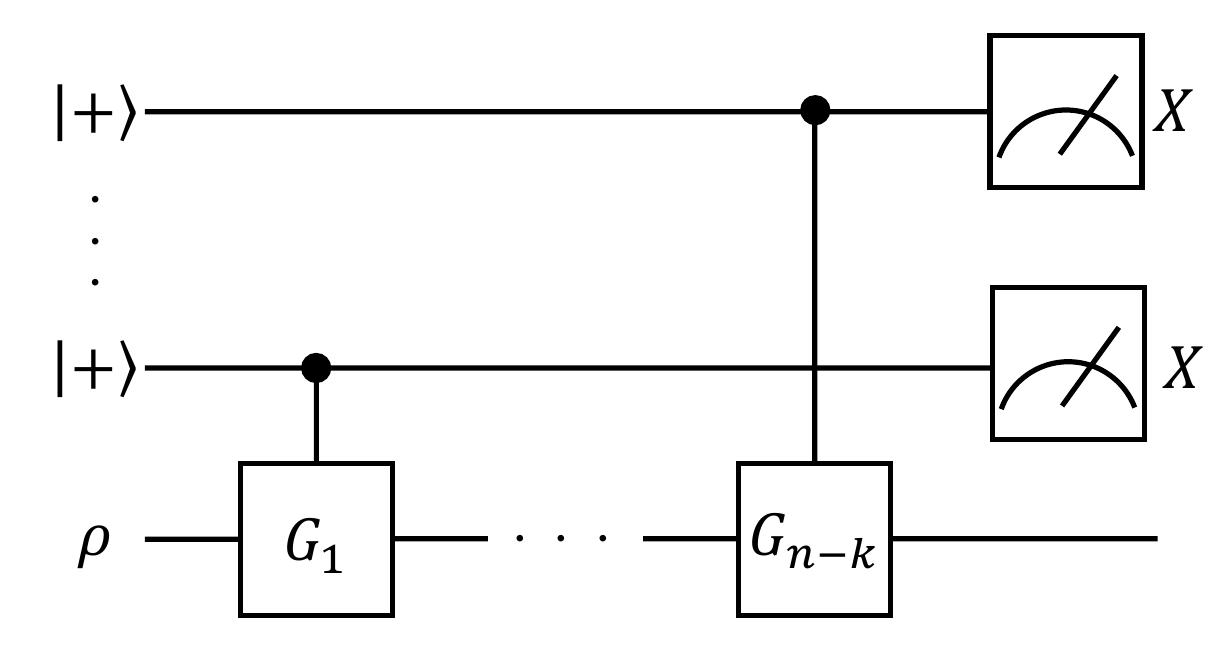}
        \caption{Quantum circuit for quantum error detection (QED).
        }
        \label{fig_QED_circuit}
    \end{center}
\end{figure}

We can detect physical errors during quantum computation by measuring the generators $G_1,\cdots,G_{n-k}$ by using the Hadamard test circuits as shown in Fig~\ref{fig_QED_circuit}, \red{and such measurement} is called syndrome measurement.
If there exists $G_i$ such that its measurement result is $-1$, then we can determine the presence of errors during the computation.
Conversely, when the measurement results are $+1$ for all $G_i$, \textcolor{black}{we can say that} there \textcolor{black}{was} no error with a sufficiently high probability.
By continuing the computation only when the measurement results for all the generators are $+1$, we can project the noisy state $\rho = \mathcal{E}(\rho_{\mathrm{id}})$ into the code space as
\begin{eqnarray}
    \rho_{\mathrm{det}} = \frac{P\rho P}{\tr[\rho P]},
\end{eqnarray}
where $P$ is an projector to the code space $\mathcal{C}$ written as
\begin{eqnarray}
P  = \prod_{G_i\in\mathcal{G}}\frac{I + G_i}{2} = \frac{1}{2^{n-k}}\sum_{S_i\in\mathcal{S}}S_i.
\end{eqnarray}
Because the probability to measure $+1$ for all the syndrome measurements is $\tr[\rho P]$, the effect of \textcolor{black}{physical} errors acting on less than $d$ qubits can be eliminated with \textcolor{black}{$O(\tr[\rho P]^{-1})$ times more execution of quantum circuits}. \textcolor{black}{Note that stabilizer-like QEM methods work in a similar way when the spin and electron number preservation is imposed in the variational ansatz of quantum states~\cite{mcardle2019error,bonet2018low}.}

We can not only detect errors but also correct them by applying appropriate feedback operations according to the measurement results, enabling us to suppress the effect of noise without \textcolor{black}{any additional execution of quantum circuits}.
\textcolor{black}{When the measurement result for the generator $G_i$ is $s_i$, and there is no measurement error, we can correct errors by applying a recovery operation $R_{\vb*{s}}$, which is estimated from ${\vb*{s}} = (s_1, ...,s_{n-k})$ to maximize the probability of correcting erroneous quantum states to the original logical state. Since the recovery Pauli operator at least maps quantum states to a logical state, $R_{\vb*{s}}$ commutes with $G_i$ if $s_i = +1$ and anti-commutes if $s_i = -1$.
%G_i'^{(1-s_i)/2}$, where $G_i'\in\mathcal{G}_n$ satisfies $[G_i',G_j] = 0$ for all $j \neq i$ and $\qty{G_i',G_i} = 0$.
In this way, the effect of \textcolor{black}{physical} errors acting on less than $\lfloor (d-1)/2 \rfloor$ qubits can be corrected as 
\begin{eqnarray}
    \label{eq_qec}
    \rho_{\mathrm{cor}} 
    = \sum_{\vb*{s}\in\qty{-1,1}^{n-k}}E_{\vb*{s}}\rho E_{\vb*{s}},
\end{eqnarray}
where
\begin{equation}
\begin{aligned}
    E_{\vb*{s}} 
    &= R_{\vb*{s}} \prod_{i}\frac{I + s_iG_i}{2}\\
    &= P R_{\vb*{s}}.
\end{aligned}
\end{equation}
%The estimated errors are stored in the Pauli frame and used for correcting logical Pauli measurement results. Even if we consider measurement errors, we can estimate appropriate recovery Pauli operators with $d$-cycle history of measurement values.
}

\textcolor{black}{
While QEC and QED can reduce the effective error rates, they impose additional difficulties in the implementation. 
To implement QEC and QED, we need repetitive applications of Pauli measurements for all the elements in the stabilizer generator set.
Since the error rates of measurement operations are typically higher than the others~\cite{hicks2022active,gunther2021improving,arute2019quantum}, they induce large overheads on the process.
In the case of QEC, we also need to estimate recovery operations from the observed syndrome values, and error rates must be smaller than the value called code threshold for a reliable estimation.}

%Therefore, an alternative way to suppress errors is demanded for near-term devices.
% QEC and QED need stablizer measurement, but 
% 1. we need to measure all the generators
% 2. the fidleity of measurement is higher than the other operations.
% In particular, QEC needs additional ...
% 1. decoding algorithm
% 2. error rates must smaller than the threshold of QEC codes.
%For a reliable error estimation, physical error rates must be sufficiently smaller than a value called the threshold. While physical error rates are steadily suppressed, this requirement is still difficult for current technologies. In particular, the physical error rate of readout operations is still higher than the threshold.
%The other burden of QEC  is the procedure for estimating a recovery operation $R$. This procedure is not necessarily efficient but must be faster than the period of syndrome measurements. While several approximated algorithms are proposed, this degrades the reduction of logical error rates.}
%performing adaptive operations during the computations requires sophisticated hardware to be constructed, making error correction difficult to implement.
%The high error rates of the measurement are also an obstacle to detecting errors during the computation~\cite{please_cite_arute_google_supremacy_to_show_high_readout_errors

\subsection{Symmetry expansion}
\label{sec_2}
In order to combat errors on near-term devices, QEM has been developed in recent years~\cite{endo2021hybrid,cai2022quantum}.
\textcolor{black}{Symmetry expansion (SE) is one of the promising QEM methods \textcolor{black}{which} mitigates errors by virtually projecting the noisy quantum state onto the symmetric subspace without syndrome measurements~\cite{mcclean2020decoding,cai2021quantum}.}

%\textcolor{black}{Since Stabilizer codes have execess $\mathbb{Z}_2$ symmetries, we can apply SE to detect errors on Stabilizer codes.}
%\textcolor{black}{SE provides us with powerful tools to suppress the effect of errors }

%can encode logical qubits into enlarged physical qubits but performing syndrome measurements, decoding of errors, and adaptive operations during computations are difficult to implement.

Suppose that we want to estimate an expectation value of an observable $O$ for a noiseless state $\rho_{\mathrm{id}}$ from the measurement of the noisy state $\rho = \mathcal{E}(\rho_{\mathrm{id}})$.
We assume that the observable $O$ commutes with the projector $P$.
Then, we can mitigate errors by virtually projecting the noisy states onto the code space as
\begin{equation}
\begin{aligned}
    \label{eq_se}
    \tr[\rho_{\mathrm{det}}O]
    &= \frac{\tr[\rho OP]}{\tr[\rho P]} \\
    &= \frac{2^{-(n-k)}\sum_{S_i \in \mathcal{S}_i}\tr[\rho OS_i]}{2^{-(n-k)}\sum_{S_i \in \mathcal{S}_i}\tr[\rho S_i]},
\end{aligned}
\end{equation}
which can be calculated in the following way.
\begin{enumerate}
    \item For $s = 1, \cdots, N$, repeat the following operations.
    \begin{enumerate}
        \item Uniformally sample $S_i\in\mathcal{S}$.
        \item Simultaneously measure the noisy state $\rho$ for $S_i$ and $OS_i$, and record the results as $a_s$ and $b_s$.
    \end{enumerate}
    \item Calculate $a = \frac{1}{N}\sum_sa_s$ and $b = \frac{1}{N} \sum_s b_s$.
    \item Output $b/a$.
\end{enumerate}
The number of measurements needed to estimate Eq. \eqref{eq_se} \textcolor{black}{for some fixed accuracy $\varepsilon$} is known to scale as $N = O(\varepsilon^{-2} \tr[\rho P]^{-2})$. In this way, we can obtain an error-mitigated expectation value of the observable $O$, which corresponds to the virtual projection of the noisy state \red{immediately before the measurement} $\rho$ onto $\rho_{\mathrm{det}}$.

\section{Virtual quantum error detection}
\label{sec_3}
Symmetry expansion is only applicable to the state immediately before measurement~\cite{cai2021quantum,mcclean2020decoding} and state preparation for rotation symmetric bosonic codes~\cite{endo2022quantum}.
In this section, we introduce our VQED method, which allows for the computation of error-mitigated expectation values corresponding to the post-selected states after syndrome measurements during circuit execution.

\begin{figure}[t]
    \begin{center}
        \includegraphics[width=0.9\linewidth]{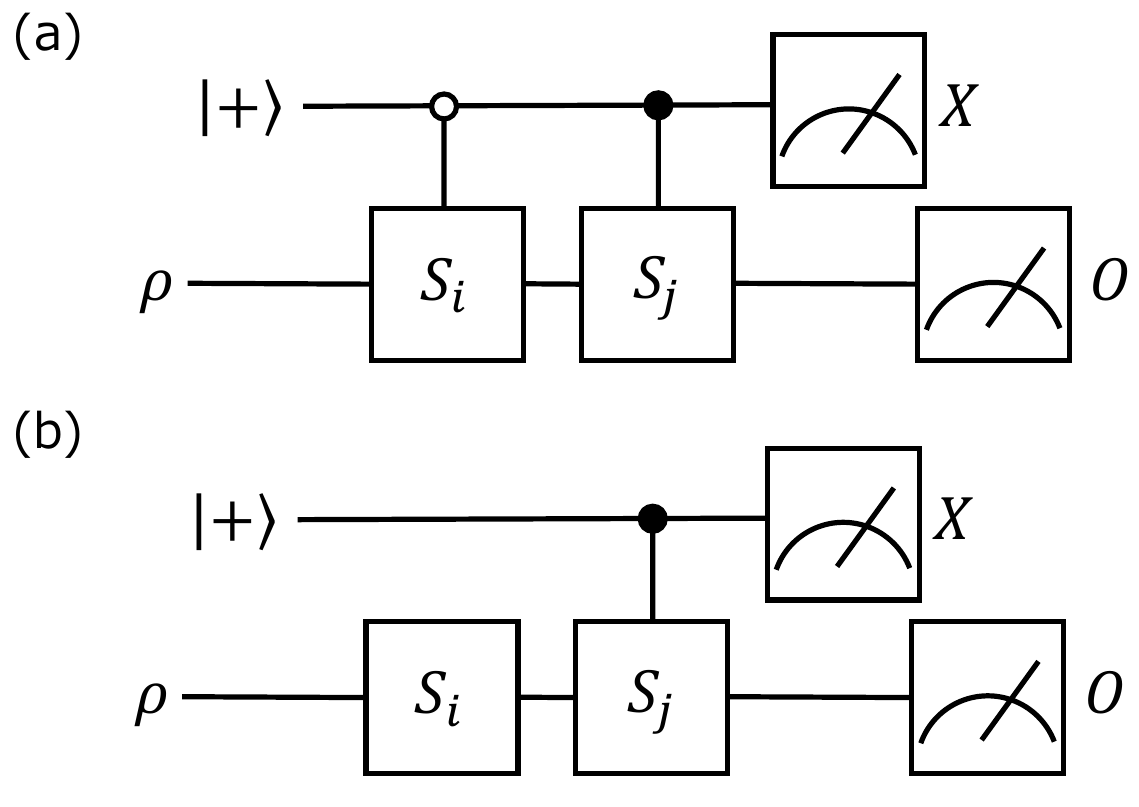}
        \caption{\redd{Quantum circuits for virtually projecting a quantum state $\rho$ into the code space. The white/black circles indicate the control operations which act for the state $0/1$. The circuit in Panel (a) utilizes two controlled-stabilizer gates, whereas the circuit in Panel (b) can virtually project quantum states using only a single controlled-stabilizer gate.}
        }
        \label{fig_vp_circuit}
    \end{center}
\end{figure}

\red{Let us first explain the way to virtually project a quantum state $\rho$ into the code space by using the circuit shown in Fig. \ref{fig_vp_circuit} (a).
We can derive the expectation value obtained through this circuit as:
\begin{eqnarray}
    \frac{1}{2}(\tr[S_i\rho S_jO] + \tr[S_j\rho S_iO]).
\end{eqnarray}
Thus, by uniformly sampling $i, j\in \qty{1,\cdots, 2^{n-k}}$ and taking the average of the distribution, we can obtain the expectation value of the projected state as
\begin{eqnarray}
    \label{eq_projected_ev}
    \tr[P\rho PO]
\end{eqnarray}
since the average of $S_i$ can be written as $\expval{S_i} = 2^{-(n-k)}\sum_{i}S_i = P$.
}

\red{We can further simplify the circuit as in Fig. \ref{fig_vp_circuit} (b) for stabilizer codes.
The expectation value obtained through this circuit is
\begin{eqnarray}
    \frac{1}{2}(\tr[S_jS_i\rho \redd{S_i}O] + \tr[S_i\rho S_iS_jO]).
\end{eqnarray}
Thus, by uniformly sampling $i, j\in \qty{1,\cdots, 2^{n-k}}$ and taking the average of the distribution, we can also obtain Eq. (\ref{eq_projected_ev}) since $PS_i =S_iP = P$ holds.
}

\red{These methods can be used to virtually detect errors in noisy quantum circuits.}
We consider a logical quantum circuit composed of a state preparation of a logical initial state $\rho_0$ followed by $L$ logical unitary gate $\mathcal{U}_l(\cdot) = U_l \cdot U_l^\dagger$ $(l = 1, \cdots, L)$, and a measurement of an observable $O$ in the hope of estimating the expectation value of $O$ for the state $\rho_{\mathrm{id}} = \mathcal{U}_L\circ\cdots \circ \mathcal{U}_1(\rho_0)$.
However, we assume that these logical quantum gates are affected by Markovian noise and that the actual gates are represented as $\mathcal{U}'_l = \mathcal{E}_l\circ\mathcal{U}_l$.
For simplicity, we ignore state preparation and measurement (SPAM) errors, but these effects can easily be reflected. When we can perform quantum error detection after each gate, we will have
\begin{equation}
\rho_{\rm det} = \frac{\rho_{\rm det}'}{\tr[\rho_{\rm det}']},
\label{Eq: deteciton}
\end{equation}
where 
\begin{equation}
\rho_{\rm det}'=\mathcal{P}\circ\mathcal{E}_L\circ\mathcal{U}_L\circ\cdots\circ\mathcal{P}\circ\mathcal{E}_1\circ\mathcal{U}_1(\rho_0).
\end{equation}
Here, we define $\mathcal{P}(\cdot) =  P\cdot P$. 

\begin{figure}[t]
    \begin{center}
        \includegraphics[width=0.9\linewidth]{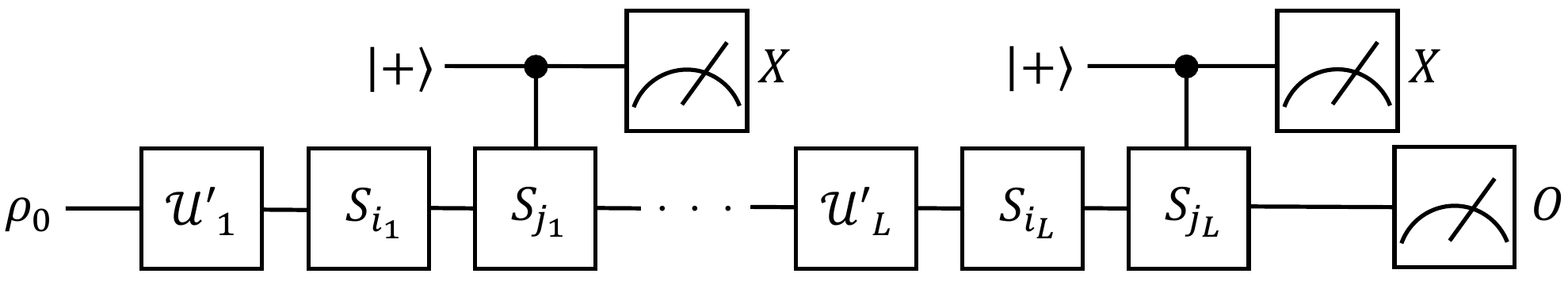}
        \caption{\red{Quantum circuit for virtual quantum error detection (VQED).}  
        }
        \label{fig_VQED_circuit}
    \end{center}
\end{figure}

In order to obtain the expectation value for $\rho_{\rm det}$ through VQED, we construct a quantum circuit represented in Fig. \ref{fig_VQED_circuit}.
This circuit allows for \red{computing} the expectation values corresponding to the error-detection circuits by \red{performing $S_{i_l}$ gate on the noisy circuit,} preparing a single qubit ancilla initialized to $\ket{+}$, coupling the ancilla qubit with the noisy circuit through controlled-$S_{j_l}$ gate, and measuring the ancilla in the $X$ bases. Note that the \red{frequency} of \red{applying these} operation for VQED can be reduced according to the noise level, although we discuss gate-wise VQED for generality.  
The state immediately before the measurement of this circuit $\rho_{\rm bf}$ reads:
\begin{eqnarray}
\begin{aligned}
    \rho_{\rm bf}&=\frac{1}{2^L}\sum_{\vb*{p}\vb*{q}}\ket{\vb*{p}}\bra{\vb*{q}}\otimes\rho_{\vb*{i}\vb*{j}}^{\vb*{p}\vb*{q}},\\
    \rho_{\vb*{i}\vb*{j}}^{\vb*{p}\vb*{q}} &= \mathcal{P}_{i_Lj_L}^{p_Lq_L}\circ\mathcal{E}_L\circ\mathcal{U}_L\circ\cdots\circ\mathcal{P}_{i_1j_1}^{p_1q_1}\circ\mathcal{E}_1\circ\mathcal{U}_1(\rho_0),
    \end{aligned}
\end{eqnarray}
where $\vb*{p}$ and $\vb*{q}$ are bitstrings of length $L$ and
\begin{eqnarray}
    \mathcal{P}_{i_lj_l}^{p_lq_l}(\cdot) = \red{S^{p_l}_{j_l}S_{i_l}\cdot S_{i_l}S^{q_l}_{j_l}}.
\end{eqnarray}
Then, the expectation value of the observable $X^{\otimes L}\otimes O$ in this state is:
\begin{eqnarray}
\begin{aligned}
    \langle X^{\otimes L} \otimes O\rangle &=\tr[\rho_{\rm bf} X^{\otimes L}\otimes O]\\
    &= \frac{1}{2^L}\sum_{\vb*{p}}\mathrm{tr}[\rho_{\vb*{i}\vb*{j}}^{\vb*{p}\vb*{p+1}}O]
\end{aligned}
\end{eqnarray}
where $\vb*{1}$ is a bit string of length $L$ whose elements are all 1.
When we uniformly sample $i_l, j_l\in \qty{1,\cdots, 2^{n-k}}$ $(1\leq l\leq L)$ and denote the expectation value under the probability distribution as $\expval{\cdot}_{\vb*{i}\vb*{j}}$, we can project the noisy state into the code space after each noisy gate as
\begin{eqnarray}
\begin{aligned}
\ev*{\rho_{\vb*{i}\vb*{j}}^{\vb*{p}\vb*{p+1}}}_{\vb*{i}\vb*{j}} &= \mathcal{P}\circ\mathcal{E}_L\circ\mathcal{U}_L\circ\cdots\circ\mathcal{P}\circ\mathcal{E}_1\circ\mathcal{U}_1(\rho_0) \\
&= \rho'_{\mathrm{det}},
\label{Eq: result1}
\end{aligned}
\end{eqnarray}
where we use 
\begin{eqnarray}
\begin{aligned}
\ev*{\mathcal{P}_{i_lj_l}^{p_lp_l+1}(\cdot)}_{i_lj_l} &= \red{\ev*{S^{p_l}_{j_l}S_{i_l}\cdot S_{i_l}S^{1-p_l}_{j_l}.}_{i_lj_l}} \\
    &= \begin{cases}
         \red{\ev*{S_{i_l}\cdot S_{i_l}S_{j_l}}_{i_lj_l}} & (p_l = 0) \\
         \red{\ev*{S_{j_l}S_{i_l}\cdot S_{i_l}}_{i_lj_l}} & (p_l=1)
        \end{cases}\\
    &= P\cdot P.
\label{Eq: result2}
\end{aligned}
\end{eqnarray}
\begin{table*}[t]
    \begin{minipage}[t]{0.33\hsize}
        (a) \red{$[[4,1,2]]$} stabilizer code
        \begin{center}
            \begin{tabular}{c|c}
                Name & Operator   \\ \hline
                $G_1$    & $XXXX$   \\
                $G_2$    & $ZZZZ$   \\
                $G_3$    & $IZZI$   \\
                $Z_L$    & $ZZII$   \\
                $X_L$    & $IXXI$   \\
            \end{tabular}
        \end{center}
    \end{minipage}
    \begin{minipage}[t]{0.33\hsize}
        (b) \red{$[[5,1,3]]$} stabilizer code
        \begin{center}
            \begin{tabular}{c|c}
                Name & Operator   \\ \hline
                $G_1$    & $XZZXI$   \\
                $G_2$    & $IXZZX$   \\
                $G_3$    & $XIXZZ$   \\
                $G_4$    & $ZXIXZ$   \\
                $Z_L$    & $ZZZZZ$   \\
                $X_L$    & $XXXXX$   \\
            \end{tabular}
        \end{center}
    \end{minipage}
    \begin{minipage}[t]{0.3\hsize}
        (c) \red{$[[7,1,3]]$} stabilizer code
        \begin{center}
            \begin{tabular}{c|c}
                Name & Operator   \\ \hline
                $G_1$    & $IIIZZZZ$   \\
                $G_2$    & $IZZIIZZ$   \\
                $G_3$    & $ZIZIZIZ$   \\
                $G_1$    & $IIIXXXX$   \\
                $G_2$    & $IXXIIXX$   \\
                $G_3$    & $XIXIXIX$   \\
                $Z_L$    & $ZZZZZZZ$   \\
                $X_L$    & $XXXXXXX$   \\
            \end{tabular}
        \end{center}
    \end{minipage}
    \caption{Generators and logical operators for (a) \red{$[[4,1,2]]$}, (b) \red{$[[5,1,3]]$}, and (c) \red{$[[7,1,3]]$} stabilizer codes.}
    \label{tabel_generators}
\end{table*}

Thus, the expectation value of the observable $O$ for the post-selected state $\rho_{\det}$ can be represented as:
\begin{equation}
\label{eq_vqed}
\begin{aligned}
    \tr[\rho_{\mathrm{det}}O] 
    = \frac{\expval{\tr\qty[\qty(\frac{1}{2^L}\sum_{\vb*{p}\vb*{q}}\ket{\vb*{p}}\bra{\vb*{q}}\otimes\rho_{\vb*{i}\vb*{j}}^{\vb*{p}\vb*{q}})X^{\otimes L}\otimes O]}_{\vb*{i}\vb*{j}}}{\expval{\tr\qty[\qty(\frac{1}{2^L}\sum_{\vb*{p}\vb*{q}}\ket{\vb*{p}}\bra{\vb*{q}}\otimes\rho_{\vb*{i}\vb*{j}}^{\vb*{p}\vb*{q}})X^{\otimes L}\otimes I]}_{\vb*{i}\vb*{j}}}.
\end{aligned}
\end{equation}
Therefore, we can perform our VQED in the noisy quantum circuit with the following procedure:
\begin{enumerate}
    \item For $s = 1,\cdots, N$, repeat the following operation.
    \begin{enumerate}
        \item Uniformly sample $i_l, j_l\in \qty{1,\cdots, 2^{n-k}}$ $(1\leq l\leq L)$.
        \item Run the circuit illustrated in Fig. \ref{fig_VQED_circuit}.
        \item Record the product of the $X$ measurement as $a_s$ and the product of $a_s$ and $O$ measurement as $b_s$.
    \end{enumerate}
    \item Calculate $a = \frac{1}{N}\sum_sa_s$ and $b = \frac{1}{N}\sum_sb_s$.
    \item Output $b/a$.
\end{enumerate}
In this way, with the sampling overhead of $N = O(\varepsilon^{-2} \tr[\rho_{\mathrm{det}}']^{-2})$, we can perform VQED \red{to virtually detect errors} that occurred during the computation with some fixed accuracy \red{(standard deviation)} $\varepsilon$. Note that while we focus on the stabilizer QEC/QED codes, our method can be straightforwardly applied to the stabilizer-based QEM method for the spin and electron number preservation~\cite{mcardle2019error,bonet2018low} for more near-term quantum hardware.

Note that our VQED protocol circumvents the syndrome measurements of stabilizer generators that need high-fidelity single-shot measurement of ancilla qubits; our method only measures expectation values of the observable. 
%Furthermore
\red{Moreover}, while quantum error detection requires measurements of $n-k$ stabilizer generators via Hadamard test circuits shown in Fig. \ref{fig_QED_circuit}, our method only necessitates \red{a single} controlled operations irrespective of the number of stabilizer generators.

\red{Furthermore, \reddd{for certain simple error
models,} the obtained expectation value is robust against noise that occurs in the ancilla qubit.
Since we calculate the expectation value of $X$ for each ancilla, the only terms of the ancilla that affect Eq. (\ref{eq_vqed}) are $\ketbra{0}{1}$ and $\ketbra{1}{0}$.
Thus, even if the single qubit depolarizing noise $\mathcal{E}_p: \rho\mapsto (1-p)\rho + pI/2$ affects each ancilla during the execution of controlled-$S_{j_l}$ gate, the numerator and the denominator of Eq. (\ref{eq_vqed}) are only multiplied by $(1-p)^{L}$.
Therefore, the value obtained through VQED remains unchanged.
Note that the circuit level noise model, where each CNOT gate and CZ gate to implement the controlled-$S_{j_l}$ gate is affected by a single qubit depolarizing noise, also does not affect the expectation value. This is because the noisy term represented as $\ketbra{0}\otimes\rho+\ketbra{1}\otimes\rho'$, where $\rho$ and $\rho'$ are the states of the system qubits where noise may be propagated, cancels out when we take the expectation value of $X$.
The same principle applies to other noise models which are not biased by Pauli $X$ or $Y$, such as local dephasing and amplitude damping noise.
\redd{We further discuss these points in Appendix \ref{sec_A1}.}
We also want to mention that},
by combining the readout error mitigation method~\cite{maciejewski2020mitigation, bravyi2021mitigating} with our method for the ancilla qubits, we can perform high-fidelity virtual projection onto the code space even \red{under the existence of measurement errors.}
%Lastly, by combining the readout error mitigation method with our method for the ancilla qubit, we can perform high-fidelity virtual projection onto the code space even when we use noisy ancilla qubits.

The disadvantages of VQED are that we can only obtain the error-mitigated expectation values, not the quantum state itself as well as quadratically worse sampling cost for the projection probability $\tr[\rho'_{\rm det}]$. While sampling costs can only be overcome by increased parallelization, lightweight quantum phase estimation algorithms only employing expectation values are proposed~\cite{suzuki2022quantum,lin2022heisenberg,wan2022randomized,zhang2022computing} in addition to the fact that most of NISQ algorithms use expectation values. 
\red{Our VQED methods can be used in such algorithms.}

\section{Virtual implementation of quantum error correction}
\begin{figure*}[t]
    \begin{center}
        \resizebox{0.99\hsize}{!}{\includegraphics{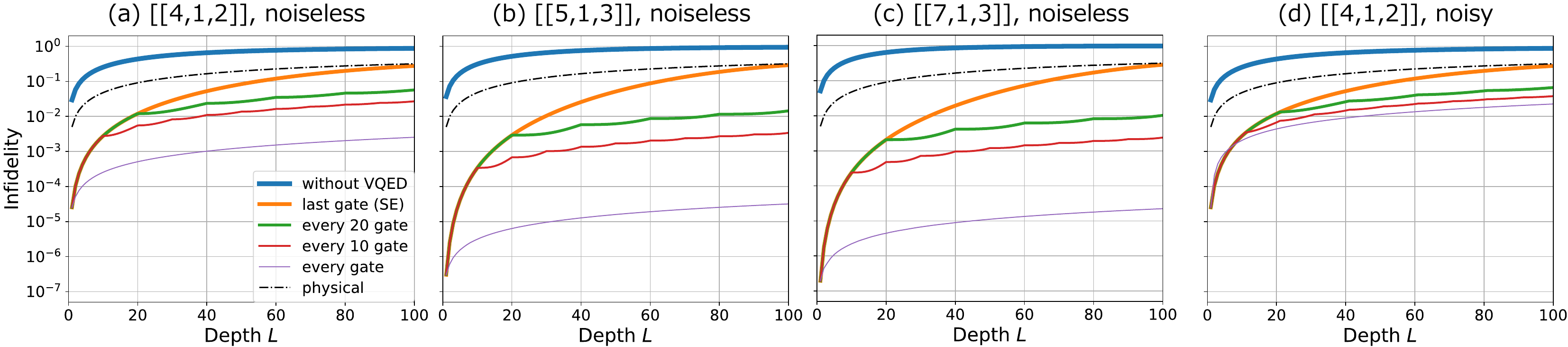}}
        \caption{Depth $L$ dependence of infidelity $1-\mathrm{tr}[\rho_{\mathrm{det}}\ketbra{\bar{\Psi}}] = 1-\ev*{\rho_{\mathrm{det}}}{\bar{\Psi}}$ between the output state of the noisy circuit with VQED $\rho_{\mathrm{det}}$ and the noiseless circuit $\ket*{\bar{\Psi}}$ for (a) \red{and (d)}: \red{$[[4,1,2]]$}, (b): \red{$[[5,1,3]]$}, and (c): \red{$[[7,1,3]]$} stabilizer codes. 
        \red{Panels (a)-(c) denotes the results when the controlled-stabilizer gates are noiseless and (d) denotes the results when the VQED gadgets are affected by local depolarizing noise.}
        The ``without \red{VQED}'' line represents infidelity when we did not perform VQED. The ``last gate \red{(SE)}'' line represents infidelity when we perform VQED only before the measurement\red{, which is just a normal SE,} as in Refs. \cite{mcclean2020decoding,cai2021quantum}. The ``every 20 gates'' and the ``every 10 gates'' lines represent infidelity when we perform VQED after every 20 and 10 gates. The ``every gate'' line represents infidelity when we perform VQED after every gate. The ``physical'' line represents the infidelity of a single physical qubit without encoding.
        }
        \label{fig_infidelity}
    \end{center}
\end{figure*}

\begin{figure*}[t]
    \begin{center}
        \resizebox{0.99\hsize}{!}{\includegraphics{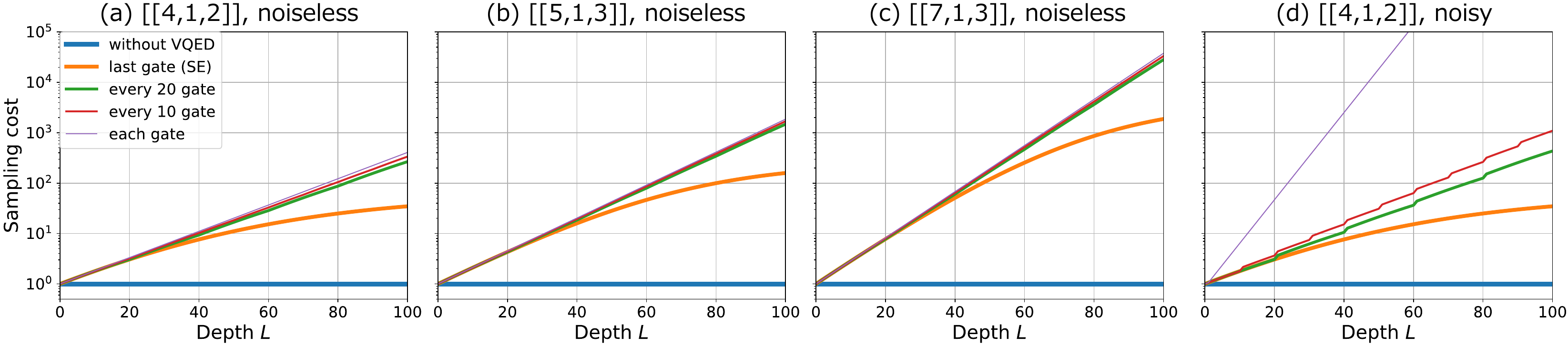}}
        \caption{Scaling of the sampling cost $\mathrm{\tr}[\rho'_{\mathrm{det}}]^{-2}$ with respect to the depth $L$ of the quantum circuit for (a) \red{and (d)}: \red{$[[4,1,2]]$}, (b): \red{$[[5,1,3]]$}, and (c): \red{$[[7,1,3]]$} stabilizer codes.
        \red{Panels (a)-(c) denotes the results when the controlled-stabilizer gates are noiseless and (d) denotes the results when the VQED gadgets are affected by local depolarizing noise.}
        The ``without \red{VQED}'' line represents \red{sampling cost} when we did not perform VQED. The ``last gate \red{(SE)}'' line represents \red{sampling cost} when we perform VQED only before the measurement\red{, which is just a normal SE,} as in Refs. \cite{mcclean2020decoding,cai2021quantum}. The ``every 20 gates'' and the ``every 10 gates'' lines represent the sampling cost when we perform VQED after every 20 and 10 gates. The ``every gate'' line represents the sampling cost when we perform VQED after every gate.
        }
        \label{fig_cost}
    \end{center}
\end{figure*}
We also discuss how to perform quantum error correction virtually without any syndrome measurements and feedback operations.
\textcolor{black}{The main idea is that the error-corrected state as in Eq. (\ref{eq_qec}) can be also written as
\begin{eqnarray}
    %\rho_{\mathrm{cor}} \\
    %= \sum_{\vb*{s}\in\qty{-1,1}^{n-k}}\mathcal{P}\qty(\prod_{i} G_i'^{(1-s_i)/2}\rho\prod_{i} G_i'^{(1-s_i)/2}).
\rho_{\mathrm{cor}} = \sum_{\vb*{s}\in\qty{-1,1}^{n-k}}\mathcal{P}\qty(R_{\vb*{s}} \rho R_{\vb*{s}}).
\end{eqnarray}
Thus, we can virtually correct errors by uniformly sampling $s_1,\cdots, s_{n-k}\in\qty{+1, -1}$, apply\textcolor{black}{ing} $R_{\vb*{s}}$ to the noisy state, virtually project\textcolor{black}{ing} the state into the codes pace using the way mentioned above, and multiplying the result by $2^{n-k}$.}
However, the sampling cost of this method scales as $2^{2(n-k)}$, which grows exponentially with the number of redundant \red{qubits}.
We may decrease the cost by limiting the scope of the sum.
Let $B \subset \qty{-1, 1}^n$ be a subset of \textcolor{black}{highly probable} measurement results such as the measurement results when an error did not occur or occurred \textcolor{black}{only once}.
\textcolor{black}{Then, we may approximate the error-corrected state as
\begin{eqnarray}
\rho_{\mathrm{cor'}} = \frac{1}{\sum_{\vb*{s}\in B} p_{\vb*{s}}}\sum_{\vb*{s}\in B}\mathcal{P}\qty( R_{\vb*{s}} \rho R_{\vb*{s}})
%    \rho_{\mathrm{cor'}}\\
%    = \frac{1}{\sum_{\vb*{s}\in B} p_{\vb*{s}}}\sum_{\vb*{s}\in B}\mathcal{P}\qty(\prod_{i} G_i'^{(1-s_i)/2}\rho\prod_{i} G_i'^{(1-s_i)/2})
\end{eqnarray}
where $p_{\vb*{s}} = \mathrm{tr}[P R_{\vb*{s}} \rho R_{\vb*{s}}]$ represents the probability of obtaining $\vb*{s} $ at the syndrome measurement.}
However, the sampling cost of virtually calculating this state is $|B|^2(1/\sum_{\vb*{s}\in B}p_{\vb*{s}})^2$, which is still significantly higher than just performing VQED with $B = \qty{1}^n$.
Furthermore, while error detection can detect errors of at most $d$ qubits, error correction can only correct errors of at most $\lfloor (d-1)/2 \rfloor$ qubits.
This means that even the accuracy of this virtual implementation of QEC is generally worse than VQED.
Even though these methods may be more effective than VQED in the case where the noise maps the state in the code space outside of it with high probability, finding practical scenarios to utilize these methods is left as our future work.

\section{Numerical simulation}
In this section, we numerically evaluate the performance of our method for \red{$[[4,1,2]]$}, \red{$[[5,1,3]]$}, and \red{$[[7,1,3]]$} stabilizer codes~\cite{grassl1997codes,laflamme1996perfect,steane1996error}.
The generators and logical operators of these codes are shown in Tabel \ref{tabel_generators}.
\red{Similar to the numerical calculation presented in the previous study on SE~\cite{mcclean2020decoding}, we initialize the state $\rho_0$ as the logical state $\ket{0}_L$, and the unitary gate $U_l$ is randomly chosen from a set of transversal single-qubit gates~\cite{gottesman2016quantum, gottesman1997stabilizer}.}
\redd{We specify the set of transversal single-qubit gates we use in Appendix \ref{sec_A2}.}
%We set the initial state $\rho_0$ to be the logical $0$ state $\ket{0}_L$ and the unitary gate $U_l = e^{i\theta_iZ_L}e^{i\theta'_iX_L}$ where $\theta_l$, $\theta_l'$ is uniformly sampled from $[0, 2\pi)$.
We assume that the local depolarizing noise $\mathcal{E} = \mathcal{E}_{p}^{\otimes n}$ ($\mathcal{E}_{p}(\rho)=(1-p)\rho+pI/2$) disturb the circuit with noise strength $p = 0.01$ each after the gate.
We numerically calculate the depth $L$ dependence of the infidelity $1-\expval*{\rho_{\mathrm{det}}}{\bar{\Psi}}$ between the output state of the noisy circuit $\rho_{\mathrm{det}}$ and the noiseless circuit $\ket*{\bar{\Psi}}$, and the scaling of the sampling cost $\mathrm{\tr}[\rho'_{\mathrm{det}}]^{-2}$ by using QuTiP~\cite{johansson2012qutip}.

Our results are shown in Fig. \ref{fig_infidelity} and Fig. \ref{fig_cost}.
As shown in Fig. \ref{fig_infidelity}, we can reduce the infidelity \red{using} VQED compared to a single physical qubit \red{affected by the error $\mathcal{E}_p$} without encoding.
\textcolor{black}{Furthermore, frequent application of VQED during the circuit execution prevents the noisy state to be highly mixed on the code space. This allows us to suppress logical errors that cannot be mitigated by the conventional SE performed only on the state immediately before the measurement~\cite{cai2021quantum,mcclean2020decoding}.}
We also find that we can reduce infidelity without performing error detection after every gate: we can sufficiently mitigate errors simply by performing
\red{VQED}
%error detection
after every fixed number of gates.
This fact can be useful when the measurement time is much longer than the gate execution time.
By comparing infidelity among different codes, we can say that infidelity becomes smaller as the code distance gets larger.
%For the same code distance, the accuracy of calculation depends on how often VQED is performed.

Fig. \ref{fig_cost} shows that the sampling cost increases exponentially with the circuit depth $L$ when we perform VQED frequently.
This scaling can be roughly considered to be given by the square inverse of the probability that a state in the code space \red{remains in the} code space when the noise is applied; thus we can say
$\red{N\sim}\tr[\rho_{\rm det}']^{-2} = O((1-\frac{3}{4}p)^{-2nL})$.
%Thus, the cost grows exponentially with the number of qubits $n$ we use to encode.
\red{It is noteworthy that the sampling cost is not significantly influenced by the frequency of VQED, even when comparing the cases of the conventional SE and VQED applied each after the gate (See the numerical results up to the depth $L=40$, for example). 
Meanwhile, it may appear that the sampling cost of SE approaches a constant value for large $L$.
However, this is because the accumulated errors increase and the noisy state approaches a completely mixed state, and thus the sampling cost converges to $N\sim \mathrm{tr}[I/2^nP] = \frac{1}{2^{n-k}}$.}
%when VQED is performed only before the measurement, the sampling cost approaches a constant value.
%This is because the noisy state approaches the completely mixed state, so the probability of \textcolor{black}{not} detecting an error approaches $\red{N\sim} \mathrm{tr}[I/2^nP] = \frac{1}{2^{n-k}}$.

In Fig. \ref{fig_infidelity} (d) and Fig. \ref{fig_cost} (d), we also present the performance of VQED \redd{for $[[4, 1, 2]]$ stabilizer code} when the VQED gadgets $S_{i_l}$ gate and controlled-$S_{j_l}$ gate in Fig. \ref{fig_VQED_circuit} are \redd{each} affected by the local depolarizing noise \redd{$\mathcal{E}_{p}^{\otimes n}$ and $\mathcal{E}_{p}^{\otimes (n+1)}$} with the same error rate $p=0.01$.
\redd{See Appendix \ref{sec_A3} for the results for $[[5, 1, 3]]$ and $[[7, 1, 3]]$ stabilizer codes.}
In the previous section, we have mentioned that a wide class of realistic noise that occurred in the ancilla does not affect the accuracy of VQED.
By comparing Fig. \ref{fig_infidelity} (a) and Fig. \ref{fig_infidelity} (d), we can also say that even in the presence of additional noise acting on the system qubits, VQED still significantly outperforms the unmitigated results and the conventional SE. Meanwhile, we can see from Fig. \ref{fig_cost} (a) and Fig. \ref{fig_cost} (d) that the sampling cost increases when we assume that the noise affects the VQED gadget.
This is mainly because the probability that a state in the code space remains in the space decreases due to the additional noise acting on the system qubits, and the denominator in Eq. (\ref{eq_vqed}) is multiplied by the factor of $(1-p)^{L}$.
Thus, the sampling cost scales as $N\sim O((1-\frac{3}{4}p)^{-6nL}(1-p)^{-2L})$.
However, these effects can be circumvented by reducing the frequency of applying the operations of VQED.

From the above, we can say that our method can be used effectively by adjusting the code distance or frequency of VQED according to the hardware constraints, the desired accuracy, or the allowable sampling cost. 
%We also remark that even when the noisy state is projected onto the code space with the conventional symmetry expansion immediately before measurement, the error-mitigated state is still highly mixed on the code space due to logical errors. Therefore, frequent application of VQED is also important for suppressing logical errors to obtain the expectation values corresponding to a pure state. 

\section{Discussion}
We propose virtual quantum error detection (VQED) so that the computation errors during the circuit execution can be flexibly suppressed by using additional two-qubit operations and measurements in the $X$ basis. We verify in the numerical simulations that our virtual quantum error detection protocol allows for the realization of significantly higher-fidelity calculation of expectation values, compared with the conventional symmetry expansion method, at the cost of sampling costs. We also discuss the virtual implementation of quantum error correction; however, even though the fidelity of the quantum state after quantum error correction is generally lower than that for quantum error detection, the sampling cost of virtual implementation of quantum error correction becomes larger than VQED. 

Although we mainly discuss the stabilizer codes based on the Pauli group, we can apply our method to other types of codes such as rotation symmetric bosonic codes (RSBCs)~\cite{grimsmo2020quantum} as well. In Ref. \cite{endo2022quantum}, symmetry expansion in RSBCs is proposed, but it is restricted to state preparation and immediately before measurement. By considering the rotation symmetry operators rather than Pauli symmetries, we can also perform virtual quantum error detection for RSBCs, which is a significant generalization of Ref. \cite{endo2022quantum}. In this case, we need controlled-rotation gates, which is implemented by the dispersive interactions~\cite{blais2021circuit} between the resonator and the ancilla qubit.  

%Also, even after the application of VQED, there remains a finite logical error, e.g., due to the limitations of code distances. If we can characterize such logical errors, we can invert them with probabilistic error cancellation~\cite{temme2017error,endo2018practical} as discussed in Refs. \cite{suzuki2022quantum,xiong2020sampling,piveteau2021error,lostaglio2021error}. 

Even after the application of VQED, there remains a finite logical error, e.g., due to the limitations of code distances. Therefore, the efficient combination of VQED with other QEM methods, e.g, purification-based QEM~\cite{huggins2021virtual,koczor2021exponential,huo2022dual,yoshioka2022generalized,seif2023shadow,o2021error}, will also be an important research direction for the realization of even more accurate quantum computing. Also, because VQED can be regarded as a QEM method implemented on the code space, the relationship between VQED and other hybrid QEM/QEC methods are worth exploring~\cite{suzuki2022quantum,xiong2020sampling,piveteau2021error,lostaglio2021error}. 

%\red{While we can repeatedly use one ancilla qubit for VQED during the computation, we can assign different ones for each logical qubit. We can flexibly choose the arrangement of the ancilla qubits according to the constraints of the experiment. For example, for trapped ion systems, all-to-all connectivity is a relatively reasonable assumption; therefore, one ancilla qubits may be iteratively used for all logical qubits. Meanwhile, superconducting qubit devices have restricted connectivity, and it may be better to assign different ancilla qubits to each logical qubit.}

%\red{Experimental implementation of VQED is also an important direction for future work. The VQED circuit we proposed in Fig. \ref{fig_VQED_circuit} requires connectivity between the ancilla qubit and all the system qubits, and we generally need swap operations to implement controlled-stabilizer operations in the case of restricted connectivity, e.g., current superconducting hardware. 
%We need to explore more hardware-friendly ways to implement VQED and demonstrate its effectiveness in actual hardware.
%Hardware like trapped ions, where all-to-all connectivity is more natural, may be suitable for the implementation.
%}
\red{Experimental implementation of VQED is also an important direction for future work. 
The VQED circuit we propose in Fig. \ref{fig_VQED_circuit} requires connectivity between the ancilla qubit and all the qubits constructing logical qubits, and we generally need swap operations to implement controlled-stabilizer operations in the case of restricted connectivity, e.g., current superconducting hardware.
However, we can relax the requirement by assigning different ancilla qubits for each logical qubit and flexibly choosing the arrangement of the ancilla qubits according to the constraints of the experiment.
For example, for trapped ion systems, all-to-all connectivity is a relatively reasonable assumption~\cite{figgatt2019parallel}; therefore, one ancilla qubits may be iteratively used for all logical qubits.
Meanwhile, superconducting qubit devices have restricted connectivity~\cite{arute2019quantum}, and it may be better to assign different ancilla qubits to each logical qubit.
}

Finally, information-theoretic analysis of QEM is one of the intensively studied topic~\cite{takagi2021optimal,takagi2022fundamental,takagi2022universal,tsubouchi2022universal,hakoshima2021relationship,quek2022exponentially}. As far as we know, symmetries of the system is not explicitly considered in these works while our work shows that they can play a crucial role for QEM. Therefore, the construction of an information-theoretic analysis of QEM incorporating the symmetries may \textcolor{black}{shed light on} e.g., the characterization cost of the noise model for performing QEM.  

\section*{Acknowledgments}
This work is supported by PRESTO, JST, Grant No.\,JPMJPR1916, JPMJPR2114, JPMJPR2119; CREST, JST, Grant No.\,JPMJCR1771; MEXT Q-LEAP Grant No.\,JPMXS0120319794 and JPMXS0118068682, JST Moonshot R\&D, Grant No.\,JPMJMS2061,  COI-NEXT program Grant No. JPMJPF2221,
\reddd{JST CREST Grant No. JPMJCR23I4,  and JST ERATO Grant Number  JPMJER2302}.
\textcolor{black}{K.T. is supported by Worldleading Innovative Graduate Study Program for Materials Research, Industry, and Technology (MERITWINGS) of the University of Tokyo.}
\bibliographystyle{apsrev4-1}
\bibliography{bib}

%merlin.mbs apsrev4-1.bst 2010-07-25 4.21a (PWD, AO, DPC) hacked
%Control: key (0)
%Control: author (72) initials jnrlst
%Control: editor formatted (1) identically to author
%Control: production of article title (-1) disabled
%Control: page (0) single
%Control: year (1) truncated
%Control: production of eprint (0) enabled
\begin{thebibliography}{56}%
\makeatletter
\providecommand \@ifxundefined [1]{%
 \@ifx{#1\undefined}
}%
\providecommand \@ifnum [1]{%
 \ifnum #1\expandafter \@firstoftwo
 \else \expandafter \@secondoftwo
 \fi
}%
\providecommand \@ifx [1]{%
 \ifx #1\expandafter \@firstoftwo
 \else \expandafter \@secondoftwo
 \fi
}%
\providecommand \natexlab [1]{#1}%
\providecommand \enquote  [1]{``#1''}%
\providecommand \bibnamefont  [1]{#1}%
\providecommand \bibfnamefont [1]{#1}%
\providecommand \citenamefont [1]{#1}%
\providecommand \href@noop [0]{\@secondoftwo}%
\providecommand \href [0]{\begingroup \@sanitize@url \@href}%
\providecommand \@href[1]{\@@startlink{#1}\@@href}%
\providecommand \@@href[1]{\endgroup#1\@@endlink}%
\providecommand \@sanitize@url [0]{\catcode `\\12\catcode `\$12\catcode
  `\&12\catcode `\#12\catcode `\^12\catcode `\_12\catcode `\%12\relax}%
\providecommand \@@startlink[1]{}%
\providecommand \@@endlink[0]{}%
\providecommand \url  [0]{\begingroup\@sanitize@url \@url }%
\providecommand \@url [1]{\endgroup\@href {#1}{\urlprefix }}%
\providecommand \urlprefix  [0]{URL }%
\providecommand \Eprint [0]{\href }%
\providecommand \doibase [0]{http://dx.doi.org/}%
\providecommand \selectlanguage [0]{\@gobble}%
\providecommand \bibinfo  [0]{\@secondoftwo}%
\providecommand \bibfield  [0]{\@secondoftwo}%
\providecommand \translation [1]{[#1]}%
\providecommand \BibitemOpen [0]{}%
\providecommand \bibitemStop [0]{}%
\providecommand \bibitemNoStop [0]{.\EOS\space}%
\providecommand \EOS [0]{\spacefactor3000\relax}%
\providecommand \BibitemShut  [1]{\csname bibitem#1\endcsname}%
\let\auto@bib@innerbib\@empty
%</preamble>
\bibitem [{\citenamefont {Preskill}(2018)}]{preskill2018quantum}%
  \BibitemOpen
  \bibfield  {author} {\bibinfo {author} {\bibfnamefont {J.}~\bibnamefont
  {Preskill}},\ }\href@noop {} {\bibfield  {journal} {\bibinfo  {journal}
  {Quantum}\ }\textbf {\bibinfo {volume} {2}},\ \bibinfo {pages} {79} (\bibinfo
  {year} {2018})}\BibitemShut {NoStop}%
\bibitem [{\citenamefont {McArdle}\ \emph {et~al.}(2020)\citenamefont
  {McArdle}, \citenamefont {Endo}, \citenamefont {Aspuru-Guzik}, \citenamefont
  {Benjamin},\ and\ \citenamefont {Yuan}}]{mcardle2020quantum}%
  \BibitemOpen
  \bibfield  {author} {\bibinfo {author} {\bibfnamefont {S.}~\bibnamefont
  {McArdle}}, \bibinfo {author} {\bibfnamefont {S.}~\bibnamefont {Endo}},
  \bibinfo {author} {\bibfnamefont {A.}~\bibnamefont {Aspuru-Guzik}}, \bibinfo
  {author} {\bibfnamefont {S.~C.}\ \bibnamefont {Benjamin}}, \ and\ \bibinfo
  {author} {\bibfnamefont {X.}~\bibnamefont {Yuan}},\ }\href@noop {} {\bibfield
   {journal} {\bibinfo  {journal} {Reviews of Modern Physics}\ }\textbf
  {\bibinfo {volume} {92}},\ \bibinfo {pages} {015003} (\bibinfo {year}
  {2020})}\BibitemShut {NoStop}%
\bibitem [{\citenamefont {Cerezo}\ \emph {et~al.}(2021)\citenamefont {Cerezo},
  \citenamefont {Arrasmith}, \citenamefont {Babbush}, \citenamefont {Benjamin},
  \citenamefont {Endo}, \citenamefont {Fujii}, \citenamefont {McClean},
  \citenamefont {Mitarai}, \citenamefont {Yuan}, \citenamefont {Cincio} \emph
  {et~al.}}]{cerezo2021variational}%
  \BibitemOpen
  \bibfield  {author} {\bibinfo {author} {\bibfnamefont {M.}~\bibnamefont
  {Cerezo}}, \bibinfo {author} {\bibfnamefont {A.}~\bibnamefont {Arrasmith}},
  \bibinfo {author} {\bibfnamefont {R.}~\bibnamefont {Babbush}}, \bibinfo
  {author} {\bibfnamefont {S.~C.}\ \bibnamefont {Benjamin}}, \bibinfo {author}
  {\bibfnamefont {S.}~\bibnamefont {Endo}}, \bibinfo {author} {\bibfnamefont
  {K.}~\bibnamefont {Fujii}}, \bibinfo {author} {\bibfnamefont {J.~R.}\
  \bibnamefont {McClean}}, \bibinfo {author} {\bibfnamefont {K.}~\bibnamefont
  {Mitarai}}, \bibinfo {author} {\bibfnamefont {X.}~\bibnamefont {Yuan}},
  \bibinfo {author} {\bibfnamefont {L.}~\bibnamefont {Cincio}},  \emph
  {et~al.},\ }\href@noop {} {\bibfield  {journal} {\bibinfo  {journal} {Nature
  Reviews Physics}\ }\textbf {\bibinfo {volume} {3}},\ \bibinfo {pages} {625}
  (\bibinfo {year} {2021})}\BibitemShut {NoStop}%
\bibitem [{\citenamefont {Tilly}\ \emph {et~al.}(2022)\citenamefont {Tilly},
  \citenamefont {Chen}, \citenamefont {Cao}, \citenamefont {Picozzi},
  \citenamefont {Setia}, \citenamefont {Li}, \citenamefont {Grant},
  \citenamefont {Wossnig}, \citenamefont {Rungger}, \citenamefont {Booth} \emph
  {et~al.}}]{tilly2022variational}%
  \BibitemOpen
  \bibfield  {author} {\bibinfo {author} {\bibfnamefont {J.}~\bibnamefont
  {Tilly}}, \bibinfo {author} {\bibfnamefont {H.}~\bibnamefont {Chen}},
  \bibinfo {author} {\bibfnamefont {S.}~\bibnamefont {Cao}}, \bibinfo {author}
  {\bibfnamefont {D.}~\bibnamefont {Picozzi}}, \bibinfo {author} {\bibfnamefont
  {K.}~\bibnamefont {Setia}}, \bibinfo {author} {\bibfnamefont
  {Y.}~\bibnamefont {Li}}, \bibinfo {author} {\bibfnamefont {E.}~\bibnamefont
  {Grant}}, \bibinfo {author} {\bibfnamefont {L.}~\bibnamefont {Wossnig}},
  \bibinfo {author} {\bibfnamefont {I.}~\bibnamefont {Rungger}}, \bibinfo
  {author} {\bibfnamefont {G.~H.}\ \bibnamefont {Booth}},  \emph {et~al.},\
  }\href@noop {} {\bibfield  {journal} {\bibinfo  {journal} {Physics Reports}\
  }\textbf {\bibinfo {volume} {986}},\ \bibinfo {pages} {1} (\bibinfo {year}
  {2022})}\BibitemShut {NoStop}%
\bibitem [{\citenamefont {Bharti}\ \emph {et~al.}(2021)\citenamefont {Bharti},
  \citenamefont {Cervera-Lierta}, \citenamefont {Kyaw}, \citenamefont {Haug},
  \citenamefont {Alperin-Lea}, \citenamefont {Anand}, \citenamefont {Degroote},
  \citenamefont {Heimonen}, \citenamefont {Kottmann}, \citenamefont {Menke}
  \emph {et~al.}}]{bharti2021noisy}%
  \BibitemOpen
  \bibfield  {author} {\bibinfo {author} {\bibfnamefont {K.}~\bibnamefont
  {Bharti}}, \bibinfo {author} {\bibfnamefont {A.}~\bibnamefont
  {Cervera-Lierta}}, \bibinfo {author} {\bibfnamefont {T.~H.}\ \bibnamefont
  {Kyaw}}, \bibinfo {author} {\bibfnamefont {T.}~\bibnamefont {Haug}}, \bibinfo
  {author} {\bibfnamefont {S.}~\bibnamefont {Alperin-Lea}}, \bibinfo {author}
  {\bibfnamefont {A.}~\bibnamefont {Anand}}, \bibinfo {author} {\bibfnamefont
  {M.}~\bibnamefont {Degroote}}, \bibinfo {author} {\bibfnamefont
  {H.}~\bibnamefont {Heimonen}}, \bibinfo {author} {\bibfnamefont {J.~S.}\
  \bibnamefont {Kottmann}}, \bibinfo {author} {\bibfnamefont {T.}~\bibnamefont
  {Menke}},  \emph {et~al.},\ }\href@noop {} {\bibfield  {journal} {\bibinfo
  {journal} {arXiv preprint arXiv:2101.08448}\ } (\bibinfo {year}
  {2021})}\BibitemShut {NoStop}%
\bibitem [{\citenamefont {Arute}\ \emph {et~al.}(2019)\citenamefont {Arute},
  \citenamefont {Arya}, \citenamefont {Babbush}, \citenamefont {Bacon},
  \citenamefont {Bardin}, \citenamefont {Barends}, \citenamefont {Biswas},
  \citenamefont {Boixo}, \citenamefont {Brandao}, \citenamefont {Buell} \emph
  {et~al.}}]{arute2019quantum}%
  \BibitemOpen
  \bibfield  {author} {\bibinfo {author} {\bibfnamefont {F.}~\bibnamefont
  {Arute}}, \bibinfo {author} {\bibfnamefont {K.}~\bibnamefont {Arya}},
  \bibinfo {author} {\bibfnamefont {R.}~\bibnamefont {Babbush}}, \bibinfo
  {author} {\bibfnamefont {D.}~\bibnamefont {Bacon}}, \bibinfo {author}
  {\bibfnamefont {J.~C.}\ \bibnamefont {Bardin}}, \bibinfo {author}
  {\bibfnamefont {R.}~\bibnamefont {Barends}}, \bibinfo {author} {\bibfnamefont
  {R.}~\bibnamefont {Biswas}}, \bibinfo {author} {\bibfnamefont
  {S.}~\bibnamefont {Boixo}}, \bibinfo {author} {\bibfnamefont {F.~G.}\
  \bibnamefont {Brandao}}, \bibinfo {author} {\bibfnamefont {D.~A.}\
  \bibnamefont {Buell}},  \emph {et~al.},\ }\href@noop {} {\bibfield  {journal}
  {\bibinfo  {journal} {Nature}\ }\textbf {\bibinfo {volume} {574}},\ \bibinfo
  {pages} {505} (\bibinfo {year} {2019})}\BibitemShut {NoStop}%
\bibitem [{\citenamefont {Kandala}\ \emph {et~al.}(2017)\citenamefont
  {Kandala}, \citenamefont {Mezzacapo}, \citenamefont {Temme}, \citenamefont
  {Takita}, \citenamefont {Brink}, \citenamefont {Chow},\ and\ \citenamefont
  {Gambetta}}]{kandala2017hardware}%
  \BibitemOpen
  \bibfield  {author} {\bibinfo {author} {\bibfnamefont {A.}~\bibnamefont
  {Kandala}}, \bibinfo {author} {\bibfnamefont {A.}~\bibnamefont {Mezzacapo}},
  \bibinfo {author} {\bibfnamefont {K.}~\bibnamefont {Temme}}, \bibinfo
  {author} {\bibfnamefont {M.}~\bibnamefont {Takita}}, \bibinfo {author}
  {\bibfnamefont {M.}~\bibnamefont {Brink}}, \bibinfo {author} {\bibfnamefont
  {J.~M.}\ \bibnamefont {Chow}}, \ and\ \bibinfo {author} {\bibfnamefont
  {J.~M.}\ \bibnamefont {Gambetta}},\ }\href@noop {} {\bibfield  {journal}
  {\bibinfo  {journal} {Nature}\ }\textbf {\bibinfo {volume} {549}},\ \bibinfo
  {pages} {242} (\bibinfo {year} {2017})}\BibitemShut {NoStop}%
\bibitem [{\citenamefont {Madsen}\ \emph {et~al.}(2022)\citenamefont {Madsen},
  \citenamefont {Laudenbach}, \citenamefont {Askarani}, \citenamefont
  {Rortais}, \citenamefont {Vincent}, \citenamefont {Bulmer}, \citenamefont
  {Miatto}, \citenamefont {Neuhaus}, \citenamefont {Helt}, \citenamefont
  {Collins} \emph {et~al.}}]{madsen2022quantum}%
  \BibitemOpen
  \bibfield  {author} {\bibinfo {author} {\bibfnamefont {L.~S.}\ \bibnamefont
  {Madsen}}, \bibinfo {author} {\bibfnamefont {F.}~\bibnamefont {Laudenbach}},
  \bibinfo {author} {\bibfnamefont {M.~F.}\ \bibnamefont {Askarani}}, \bibinfo
  {author} {\bibfnamefont {F.}~\bibnamefont {Rortais}}, \bibinfo {author}
  {\bibfnamefont {T.}~\bibnamefont {Vincent}}, \bibinfo {author} {\bibfnamefont
  {J.~F.}\ \bibnamefont {Bulmer}}, \bibinfo {author} {\bibfnamefont {F.~M.}\
  \bibnamefont {Miatto}}, \bibinfo {author} {\bibfnamefont {L.}~\bibnamefont
  {Neuhaus}}, \bibinfo {author} {\bibfnamefont {L.~G.}\ \bibnamefont {Helt}},
  \bibinfo {author} {\bibfnamefont {M.~J.}\ \bibnamefont {Collins}},  \emph
  {et~al.},\ }\href@noop {} {\bibfield  {journal} {\bibinfo  {journal}
  {Nature}\ }\textbf {\bibinfo {volume} {606}},\ \bibinfo {pages} {75}
  (\bibinfo {year} {2022})}\BibitemShut {NoStop}%
\bibitem [{\citenamefont {Devitt}\ \emph {et~al.}(2013)\citenamefont {Devitt},
  \citenamefont {Munro},\ and\ \citenamefont {Nemoto}}]{devitt2013quantum}%
  \BibitemOpen
  \bibfield  {author} {\bibinfo {author} {\bibfnamefont {S.~J.}\ \bibnamefont
  {Devitt}}, \bibinfo {author} {\bibfnamefont {W.~J.}\ \bibnamefont {Munro}}, \
  and\ \bibinfo {author} {\bibfnamefont {K.}~\bibnamefont {Nemoto}},\
  }\href@noop {} {\bibfield  {journal} {\bibinfo  {journal} {Reports on
  Progress in Physics}\ }\textbf {\bibinfo {volume} {76}},\ \bibinfo {pages}
  {076001} (\bibinfo {year} {2013})}\BibitemShut {NoStop}%
\bibitem [{\citenamefont {Lidar}\ and\ \citenamefont
  {Brun}(2013)}]{lidar2013quantum}%
  \BibitemOpen
  \bibfield  {author} {\bibinfo {author} {\bibfnamefont {D.~A.}\ \bibnamefont
  {Lidar}}\ and\ \bibinfo {author} {\bibfnamefont {T.~A.}\ \bibnamefont
  {Brun}},\ }\href@noop {} {\emph {\bibinfo {title} {Quantum error
  correction}}}\ (\bibinfo  {publisher} {Cambridge university press},\ \bibinfo
  {year} {2013})\BibitemShut {NoStop}%
\bibitem [{\citenamefont {Grassl}\ \emph {et~al.}(1997)\citenamefont {Grassl},
  \citenamefont {Beth},\ and\ \citenamefont {Pellizzari}}]{grassl1997codes}%
  \BibitemOpen
  \bibfield  {author} {\bibinfo {author} {\bibfnamefont {M.}~\bibnamefont
  {Grassl}}, \bibinfo {author} {\bibfnamefont {T.}~\bibnamefont {Beth}}, \ and\
  \bibinfo {author} {\bibfnamefont {T.}~\bibnamefont {Pellizzari}},\
  }\href@noop {} {\bibfield  {journal} {\bibinfo  {journal} {Physical Review
  A}\ }\textbf {\bibinfo {volume} {56}},\ \bibinfo {pages} {33} (\bibinfo
  {year} {1997})}\BibitemShut {NoStop}%
\bibitem [{\citenamefont {Steane}(1996)}]{steane1996error}%
  \BibitemOpen
  \bibfield  {author} {\bibinfo {author} {\bibfnamefont {A.~M.}\ \bibnamefont
  {Steane}},\ }\href@noop {} {\bibfield  {journal} {\bibinfo  {journal}
  {Physical Review Letters}\ }\textbf {\bibinfo {volume} {77}},\ \bibinfo
  {pages} {793} (\bibinfo {year} {1996})}\BibitemShut {NoStop}%
\bibitem [{\citenamefont {Shor}(1995)}]{shor1995scheme}%
  \BibitemOpen
  \bibfield  {author} {\bibinfo {author} {\bibfnamefont {P.~W.}\ \bibnamefont
  {Shor}},\ }\href@noop {} {\bibfield  {journal} {\bibinfo  {journal} {Physical
  review A}\ }\textbf {\bibinfo {volume} {52}},\ \bibinfo {pages} {R2493}
  (\bibinfo {year} {1995})}\BibitemShut {NoStop}%
\bibitem [{\citenamefont {Laflamme}\ \emph {et~al.}(1996)\citenamefont
  {Laflamme}, \citenamefont {Miquel}, \citenamefont {Paz},\ and\ \citenamefont
  {Zurek}}]{laflamme1996perfect}%
  \BibitemOpen
  \bibfield  {author} {\bibinfo {author} {\bibfnamefont {R.}~\bibnamefont
  {Laflamme}}, \bibinfo {author} {\bibfnamefont {C.}~\bibnamefont {Miquel}},
  \bibinfo {author} {\bibfnamefont {J.~P.}\ \bibnamefont {Paz}}, \ and\
  \bibinfo {author} {\bibfnamefont {W.~H.}\ \bibnamefont {Zurek}},\ }\href@noop
  {} {\bibfield  {journal} {\bibinfo  {journal} {Physical Review Letters}\
  }\textbf {\bibinfo {volume} {77}},\ \bibinfo {pages} {198} (\bibinfo {year}
  {1996})}\BibitemShut {NoStop}%
\bibitem [{\citenamefont {Hicks}\ \emph {et~al.}(2022)\citenamefont {Hicks},
  \citenamefont {Kobrin}, \citenamefont {Bauer},\ and\ \citenamefont
  {Nachman}}]{hicks2022active}%
  \BibitemOpen
  \bibfield  {author} {\bibinfo {author} {\bibfnamefont {R.}~\bibnamefont
  {Hicks}}, \bibinfo {author} {\bibfnamefont {B.}~\bibnamefont {Kobrin}},
  \bibinfo {author} {\bibfnamefont {C.~W.}\ \bibnamefont {Bauer}}, \ and\
  \bibinfo {author} {\bibfnamefont {B.}~\bibnamefont {Nachman}},\ }\href@noop
  {} {\bibfield  {journal} {\bibinfo  {journal} {Physical Review A}\ }\textbf
  {\bibinfo {volume} {105}},\ \bibinfo {pages} {012419} (\bibinfo {year}
  {2022})}\BibitemShut {NoStop}%
\bibitem [{\citenamefont {G{\"u}nther}\ \emph {et~al.}(2021)\citenamefont
  {G{\"u}nther}, \citenamefont {Tacchino}, \citenamefont {Wootton},
  \citenamefont {Tavernelli},\ and\ \citenamefont
  {Barkoutsos}}]{gunther2021improving}%
  \BibitemOpen
  \bibfield  {author} {\bibinfo {author} {\bibfnamefont {J.~M.}\ \bibnamefont
  {G{\"u}nther}}, \bibinfo {author} {\bibfnamefont {F.}~\bibnamefont
  {Tacchino}}, \bibinfo {author} {\bibfnamefont {J.~R.}\ \bibnamefont
  {Wootton}}, \bibinfo {author} {\bibfnamefont {I.}~\bibnamefont {Tavernelli}},
  \ and\ \bibinfo {author} {\bibfnamefont {P.~K.}\ \bibnamefont {Barkoutsos}},\
  }\href@noop {} {\bibfield  {journal} {\bibinfo  {journal} {Quantum Science
  and Technology}\ }\textbf {\bibinfo {volume} {7}},\ \bibinfo {pages} {015009}
  (\bibinfo {year} {2021})}\BibitemShut {NoStop}%
\bibitem [{Note1()}]{Note1}%
  \BibitemOpen
  \bibinfo {note} {\textcolor {black}{In this work, we use the term
  ``single-shot measurement'' to represent the measurements that are performed
  only once, rather than repeating the measurement many times in order to
  obtain the expectation value of some observable. Note that the meaning is
  different from the term ``single-shot error correction''~\cite
  {bombin2015single}.}}\BibitemShut {Stop}%
\bibitem [{\citenamefont {Temme}\ \emph {et~al.}(2017)\citenamefont {Temme},
  \citenamefont {Bravyi},\ and\ \citenamefont {Gambetta}}]{temme2017error}%
  \BibitemOpen
  \bibfield  {author} {\bibinfo {author} {\bibfnamefont {K.}~\bibnamefont
  {Temme}}, \bibinfo {author} {\bibfnamefont {S.}~\bibnamefont {Bravyi}}, \
  and\ \bibinfo {author} {\bibfnamefont {J.~M.}\ \bibnamefont {Gambetta}},\
  }\href@noop {} {\bibfield  {journal} {\bibinfo  {journal} {Physical review
  letters}\ }\textbf {\bibinfo {volume} {119}},\ \bibinfo {pages} {180509}
  (\bibinfo {year} {2017})}\BibitemShut {NoStop}%
\bibitem [{\citenamefont {Li}\ and\ \citenamefont
  {Benjamin}(2017)}]{li2017efficient}%
  \BibitemOpen
  \bibfield  {author} {\bibinfo {author} {\bibfnamefont {Y.}~\bibnamefont
  {Li}}\ and\ \bibinfo {author} {\bibfnamefont {S.~C.}\ \bibnamefont
  {Benjamin}},\ }\href@noop {} {\bibfield  {journal} {\bibinfo  {journal}
  {Physical Review X}\ }\textbf {\bibinfo {volume} {7}},\ \bibinfo {pages}
  {021050} (\bibinfo {year} {2017})}\BibitemShut {NoStop}%
\bibitem [{\citenamefont {Endo}\ \emph {et~al.}(2018)\citenamefont {Endo},
  \citenamefont {Benjamin},\ and\ \citenamefont {Li}}]{endo2018practical}%
  \BibitemOpen
  \bibfield  {author} {\bibinfo {author} {\bibfnamefont {S.}~\bibnamefont
  {Endo}}, \bibinfo {author} {\bibfnamefont {S.~C.}\ \bibnamefont {Benjamin}},
  \ and\ \bibinfo {author} {\bibfnamefont {Y.}~\bibnamefont {Li}},\ }\href@noop
  {} {\bibfield  {journal} {\bibinfo  {journal} {Physical Review X}\ }\textbf
  {\bibinfo {volume} {8}},\ \bibinfo {pages} {031027} (\bibinfo {year}
  {2018})}\BibitemShut {NoStop}%
\bibitem [{\citenamefont {Endo}\ \emph {et~al.}(2021)\citenamefont {Endo},
  \citenamefont {Cai}, \citenamefont {Benjamin},\ and\ \citenamefont
  {Yuan}}]{endo2021hybrid}%
  \BibitemOpen
  \bibfield  {author} {\bibinfo {author} {\bibfnamefont {S.}~\bibnamefont
  {Endo}}, \bibinfo {author} {\bibfnamefont {Z.}~\bibnamefont {Cai}}, \bibinfo
  {author} {\bibfnamefont {S.~C.}\ \bibnamefont {Benjamin}}, \ and\ \bibinfo
  {author} {\bibfnamefont {X.}~\bibnamefont {Yuan}},\ }\href@noop {} {\bibfield
   {journal} {\bibinfo  {journal} {Journal of the Physical Society of Japan}\
  }\textbf {\bibinfo {volume} {90}},\ \bibinfo {pages} {032001} (\bibinfo
  {year} {2021})}\BibitemShut {NoStop}%
\bibitem [{\citenamefont {Cai}\ \emph {et~al.}(2022)\citenamefont {Cai},
  \citenamefont {Babbush}, \citenamefont {Benjamin}, \citenamefont {Endo},
  \citenamefont {Huggins}, \citenamefont {Li}, \citenamefont {McClean},\ and\
  \citenamefont {O'Brien}}]{cai2022quantum}%
  \BibitemOpen
  \bibfield  {author} {\bibinfo {author} {\bibfnamefont {Z.}~\bibnamefont
  {Cai}}, \bibinfo {author} {\bibfnamefont {R.}~\bibnamefont {Babbush}},
  \bibinfo {author} {\bibfnamefont {S.~C.}\ \bibnamefont {Benjamin}}, \bibinfo
  {author} {\bibfnamefont {S.}~\bibnamefont {Endo}}, \bibinfo {author}
  {\bibfnamefont {W.~J.}\ \bibnamefont {Huggins}}, \bibinfo {author}
  {\bibfnamefont {Y.}~\bibnamefont {Li}}, \bibinfo {author} {\bibfnamefont
  {J.~R.}\ \bibnamefont {McClean}}, \ and\ \bibinfo {author} {\bibfnamefont
  {T.~E.}\ \bibnamefont {O'Brien}},\ }\href@noop {} {\bibfield  {journal}
  {\bibinfo  {journal} {arXiv preprint arXiv:2210.00921}\ } (\bibinfo {year}
  {2022})}\BibitemShut {NoStop}%
\bibitem [{\citenamefont {Bonet-Monroig}\ \emph {et~al.}(2018)\citenamefont
  {Bonet-Monroig}, \citenamefont {Sagastizabal}, \citenamefont {Singh},\ and\
  \citenamefont {O'Brien}}]{bonet2018low}%
  \BibitemOpen
  \bibfield  {author} {\bibinfo {author} {\bibfnamefont {X.}~\bibnamefont
  {Bonet-Monroig}}, \bibinfo {author} {\bibfnamefont {R.}~\bibnamefont
  {Sagastizabal}}, \bibinfo {author} {\bibfnamefont {M.}~\bibnamefont {Singh}},
  \ and\ \bibinfo {author} {\bibfnamefont {T.}~\bibnamefont {O'Brien}},\
  }\href@noop {} {\bibfield  {journal} {\bibinfo  {journal} {Physical Review
  A}\ }\textbf {\bibinfo {volume} {98}},\ \bibinfo {pages} {062339} (\bibinfo
  {year} {2018})}\BibitemShut {NoStop}%
\bibitem [{\citenamefont {McClean}\ \emph {et~al.}(2020)\citenamefont
  {McClean}, \citenamefont {Jiang}, \citenamefont {Rubin}, \citenamefont
  {Babbush},\ and\ \citenamefont {Neven}}]{mcclean2020decoding}%
  \BibitemOpen
  \bibfield  {author} {\bibinfo {author} {\bibfnamefont {J.~R.}\ \bibnamefont
  {McClean}}, \bibinfo {author} {\bibfnamefont {Z.}~\bibnamefont {Jiang}},
  \bibinfo {author} {\bibfnamefont {N.~C.}\ \bibnamefont {Rubin}}, \bibinfo
  {author} {\bibfnamefont {R.}~\bibnamefont {Babbush}}, \ and\ \bibinfo
  {author} {\bibfnamefont {H.}~\bibnamefont {Neven}},\ }\href@noop {}
  {\bibfield  {journal} {\bibinfo  {journal} {Nature communications}\ }\textbf
  {\bibinfo {volume} {11}},\ \bibinfo {pages} {1} (\bibinfo {year}
  {2020})}\BibitemShut {NoStop}%
\bibitem [{\citenamefont {Cai}(2021)}]{cai2021quantum}%
  \BibitemOpen
  \bibfield  {author} {\bibinfo {author} {\bibfnamefont {Z.}~\bibnamefont
  {Cai}},\ }\href@noop {} {\bibfield  {journal} {\bibinfo  {journal} {Quantum}\
  }\textbf {\bibinfo {volume} {5}},\ \bibinfo {pages} {548} (\bibinfo {year}
  {2021})}\BibitemShut {NoStop}%
\bibitem [{\citenamefont {Endo}\ \emph {et~al.}(2022)\citenamefont {Endo},
  \citenamefont {Suzuki}, \citenamefont {Tsubouchi}, \citenamefont {Asaoka},
  \citenamefont {Yamamoto}, \citenamefont {Matsuzaki},\ and\ \citenamefont
  {Tokunaga}}]{endo2022quantum}%
  \BibitemOpen
  \bibfield  {author} {\bibinfo {author} {\bibfnamefont {S.}~\bibnamefont
  {Endo}}, \bibinfo {author} {\bibfnamefont {Y.}~\bibnamefont {Suzuki}},
  \bibinfo {author} {\bibfnamefont {K.}~\bibnamefont {Tsubouchi}}, \bibinfo
  {author} {\bibfnamefont {R.}~\bibnamefont {Asaoka}}, \bibinfo {author}
  {\bibfnamefont {K.}~\bibnamefont {Yamamoto}}, \bibinfo {author}
  {\bibfnamefont {Y.}~\bibnamefont {Matsuzaki}}, \ and\ \bibinfo {author}
  {\bibfnamefont {Y.}~\bibnamefont {Tokunaga}},\ }\href@noop {} {\bibfield
  {journal} {\bibinfo  {journal} {arXiv preprint arXiv:2211.06164}\ } (\bibinfo
  {year} {2022})}\BibitemShut {NoStop}%
\bibitem [{\citenamefont {Maciejewski}\ \emph {et~al.}(2020)\citenamefont
  {Maciejewski}, \citenamefont {Zimbor{\'a}s},\ and\ \citenamefont
  {Oszmaniec}}]{maciejewski2020mitigation}%
  \BibitemOpen
  \bibfield  {author} {\bibinfo {author} {\bibfnamefont {F.~B.}\ \bibnamefont
  {Maciejewski}}, \bibinfo {author} {\bibfnamefont {Z.}~\bibnamefont
  {Zimbor{\'a}s}}, \ and\ \bibinfo {author} {\bibfnamefont {M.}~\bibnamefont
  {Oszmaniec}},\ }\href@noop {} {\bibfield  {journal} {\bibinfo  {journal}
  {Quantum}\ }\textbf {\bibinfo {volume} {4}},\ \bibinfo {pages} {257}
  (\bibinfo {year} {2020})}\BibitemShut {NoStop}%
\bibitem [{\citenamefont {Bravyi}\ \emph {et~al.}(2021)\citenamefont {Bravyi},
  \citenamefont {Sheldon}, \citenamefont {Kandala}, \citenamefont {Mckay},\
  and\ \citenamefont {Gambetta}}]{bravyi2021mitigating}%
  \BibitemOpen
  \bibfield  {author} {\bibinfo {author} {\bibfnamefont {S.}~\bibnamefont
  {Bravyi}}, \bibinfo {author} {\bibfnamefont {S.}~\bibnamefont {Sheldon}},
  \bibinfo {author} {\bibfnamefont {A.}~\bibnamefont {Kandala}}, \bibinfo
  {author} {\bibfnamefont {D.~C.}\ \bibnamefont {Mckay}}, \ and\ \bibinfo
  {author} {\bibfnamefont {J.~M.}\ \bibnamefont {Gambetta}},\ }\href@noop {}
  {\bibfield  {journal} {\bibinfo  {journal} {Physical Review A}\ }\textbf
  {\bibinfo {volume} {103}},\ \bibinfo {pages} {042605} (\bibinfo {year}
  {2021})}\BibitemShut {NoStop}%
\bibitem [{\citenamefont {McArdle}\ \emph {et~al.}(2019)\citenamefont
  {McArdle}, \citenamefont {Yuan},\ and\ \citenamefont
  {Benjamin}}]{mcardle2019error}%
  \BibitemOpen
  \bibfield  {author} {\bibinfo {author} {\bibfnamefont {S.}~\bibnamefont
  {McArdle}}, \bibinfo {author} {\bibfnamefont {X.}~\bibnamefont {Yuan}}, \
  and\ \bibinfo {author} {\bibfnamefont {S.}~\bibnamefont {Benjamin}},\
  }\href@noop {} {\bibfield  {journal} {\bibinfo  {journal} {Physical review
  letters}\ }\textbf {\bibinfo {volume} {122}},\ \bibinfo {pages} {180501}
  (\bibinfo {year} {2019})}\BibitemShut {NoStop}%
\bibitem [{\citenamefont {Nielsen}\ and\ \citenamefont
  {Chuang}(2002)}]{nielsen2002quantum}%
  \BibitemOpen
  \bibfield  {author} {\bibinfo {author} {\bibfnamefont {M.~A.}\ \bibnamefont
  {Nielsen}}\ and\ \bibinfo {author} {\bibfnamefont {I.}~\bibnamefont
  {Chuang}},\ }\href@noop {} {\enquote {\bibinfo {title} {Quantum computation
  and quantum information},}\ } (\bibinfo {year} {2002})\BibitemShut {NoStop}%
\bibitem [{\citenamefont {Gottesman}(1997)}]{gottesman1997stabilizer}%
  \BibitemOpen
  \bibfield  {author} {\bibinfo {author} {\bibfnamefont {D.}~\bibnamefont
  {Gottesman}},\ }\href@noop {} {\emph {\bibinfo {title} {Stabilizer codes and
  quantum error correction}}}\ (\bibinfo  {publisher} {California Institute of
  Technology},\ \bibinfo {year} {1997})\BibitemShut {NoStop}%
\bibitem [{\citenamefont {Suzuki}\ \emph {et~al.}(2022)\citenamefont {Suzuki},
  \citenamefont {Endo}, \citenamefont {Fujii},\ and\ \citenamefont
  {Tokunaga}}]{suzuki2022quantum}%
  \BibitemOpen
  \bibfield  {author} {\bibinfo {author} {\bibfnamefont {Y.}~\bibnamefont
  {Suzuki}}, \bibinfo {author} {\bibfnamefont {S.}~\bibnamefont {Endo}},
  \bibinfo {author} {\bibfnamefont {K.}~\bibnamefont {Fujii}}, \ and\ \bibinfo
  {author} {\bibfnamefont {Y.}~\bibnamefont {Tokunaga}},\ }\href@noop {}
  {\bibfield  {journal} {\bibinfo  {journal} {PRX Quantum}\ }\textbf {\bibinfo
  {volume} {3}},\ \bibinfo {pages} {010345} (\bibinfo {year}
  {2022})}\BibitemShut {NoStop}%
\bibitem [{\citenamefont {Lin}\ and\ \citenamefont
  {Tong}(2022)}]{lin2022heisenberg}%
  \BibitemOpen
  \bibfield  {author} {\bibinfo {author} {\bibfnamefont {L.}~\bibnamefont
  {Lin}}\ and\ \bibinfo {author} {\bibfnamefont {Y.}~\bibnamefont {Tong}},\
  }\href@noop {} {\bibfield  {journal} {\bibinfo  {journal} {PRX Quantum}\
  }\textbf {\bibinfo {volume} {3}},\ \bibinfo {pages} {010318} (\bibinfo {year}
  {2022})}\BibitemShut {NoStop}%
\bibitem [{\citenamefont {Wan}\ \emph {et~al.}(2022)\citenamefont {Wan},
  \citenamefont {Berta},\ and\ \citenamefont {Campbell}}]{wan2022randomized}%
  \BibitemOpen
  \bibfield  {author} {\bibinfo {author} {\bibfnamefont {K.}~\bibnamefont
  {Wan}}, \bibinfo {author} {\bibfnamefont {M.}~\bibnamefont {Berta}}, \ and\
  \bibinfo {author} {\bibfnamefont {E.~T.}\ \bibnamefont {Campbell}},\
  }\href@noop {} {\bibfield  {journal} {\bibinfo  {journal} {Physical Review
  Letters}\ }\textbf {\bibinfo {volume} {129}},\ \bibinfo {pages} {030503}
  (\bibinfo {year} {2022})}\BibitemShut {NoStop}%
\bibitem [{\citenamefont {Zhang}\ \emph {et~al.}(2022)\citenamefont {Zhang},
  \citenamefont {Wang},\ and\ \citenamefont {Johnson}}]{zhang2022computing}%
  \BibitemOpen
  \bibfield  {author} {\bibinfo {author} {\bibfnamefont {R.}~\bibnamefont
  {Zhang}}, \bibinfo {author} {\bibfnamefont {G.}~\bibnamefont {Wang}}, \ and\
  \bibinfo {author} {\bibfnamefont {P.}~\bibnamefont {Johnson}},\ }\href@noop
  {} {\bibfield  {journal} {\bibinfo  {journal} {Quantum}\ }\textbf {\bibinfo
  {volume} {6}},\ \bibinfo {pages} {761} (\bibinfo {year} {2022})}\BibitemShut
  {NoStop}%
\bibitem [{\citenamefont {Gottesman}(2016)}]{gottesman2016quantum}%
  \BibitemOpen
  \bibfield  {author} {\bibinfo {author} {\bibfnamefont {D.}~\bibnamefont
  {Gottesman}},\ }\href@noop {} {\bibfield  {journal} {\bibinfo  {journal}
  {arXiv preprint arXiv:1610.03507}\ } (\bibinfo {year} {2016})}\BibitemShut
  {NoStop}%
\bibitem [{\citenamefont {Johansson}\ \emph {et~al.}(2012)\citenamefont
  {Johansson}, \citenamefont {Nation},\ and\ \citenamefont
  {Nori}}]{johansson2012qutip}%
  \BibitemOpen
  \bibfield  {author} {\bibinfo {author} {\bibfnamefont {J.~R.}\ \bibnamefont
  {Johansson}}, \bibinfo {author} {\bibfnamefont {P.~D.}\ \bibnamefont
  {Nation}}, \ and\ \bibinfo {author} {\bibfnamefont {F.}~\bibnamefont
  {Nori}},\ }\href@noop {} {\bibfield  {journal} {\bibinfo  {journal} {Computer
  Physics Communications}\ }\textbf {\bibinfo {volume} {183}},\ \bibinfo
  {pages} {1760} (\bibinfo {year} {2012})}\BibitemShut {NoStop}%
\bibitem [{\citenamefont {Grimsmo}\ \emph {et~al.}(2020)\citenamefont
  {Grimsmo}, \citenamefont {Combes},\ and\ \citenamefont
  {Baragiola}}]{grimsmo2020quantum}%
  \BibitemOpen
  \bibfield  {author} {\bibinfo {author} {\bibfnamefont {A.~L.}\ \bibnamefont
  {Grimsmo}}, \bibinfo {author} {\bibfnamefont {J.}~\bibnamefont {Combes}}, \
  and\ \bibinfo {author} {\bibfnamefont {B.~Q.}\ \bibnamefont {Baragiola}},\
  }\href@noop {} {\bibfield  {journal} {\bibinfo  {journal} {Physical Review
  X}\ }\textbf {\bibinfo {volume} {10}},\ \bibinfo {pages} {011058} (\bibinfo
  {year} {2020})}\BibitemShut {NoStop}%
\bibitem [{\citenamefont {Blais}\ \emph {et~al.}(2021)\citenamefont {Blais},
  \citenamefont {Grimsmo}, \citenamefont {Girvin},\ and\ \citenamefont
  {Wallraff}}]{blais2021circuit}%
  \BibitemOpen
  \bibfield  {author} {\bibinfo {author} {\bibfnamefont {A.}~\bibnamefont
  {Blais}}, \bibinfo {author} {\bibfnamefont {A.~L.}\ \bibnamefont {Grimsmo}},
  \bibinfo {author} {\bibfnamefont {S.~M.}\ \bibnamefont {Girvin}}, \ and\
  \bibinfo {author} {\bibfnamefont {A.}~\bibnamefont {Wallraff}},\ }\href@noop
  {} {\bibfield  {journal} {\bibinfo  {journal} {Reviews of Modern Physics}\
  }\textbf {\bibinfo {volume} {93}},\ \bibinfo {pages} {025005} (\bibinfo
  {year} {2021})}\BibitemShut {NoStop}%
\bibitem [{\citenamefont {Huggins}\ \emph {et~al.}(2021)\citenamefont
  {Huggins}, \citenamefont {McArdle}, \citenamefont {O’Brien}, \citenamefont
  {Lee}, \citenamefont {Rubin}, \citenamefont {Boixo}, \citenamefont {Whaley},
  \citenamefont {Babbush},\ and\ \citenamefont {McClean}}]{huggins2021virtual}%
  \BibitemOpen
  \bibfield  {author} {\bibinfo {author} {\bibfnamefont {W.~J.}\ \bibnamefont
  {Huggins}}, \bibinfo {author} {\bibfnamefont {S.}~\bibnamefont {McArdle}},
  \bibinfo {author} {\bibfnamefont {T.~E.}\ \bibnamefont {O’Brien}}, \bibinfo
  {author} {\bibfnamefont {J.}~\bibnamefont {Lee}}, \bibinfo {author}
  {\bibfnamefont {N.~C.}\ \bibnamefont {Rubin}}, \bibinfo {author}
  {\bibfnamefont {S.}~\bibnamefont {Boixo}}, \bibinfo {author} {\bibfnamefont
  {K.~B.}\ \bibnamefont {Whaley}}, \bibinfo {author} {\bibfnamefont
  {R.}~\bibnamefont {Babbush}}, \ and\ \bibinfo {author} {\bibfnamefont
  {J.~R.}\ \bibnamefont {McClean}},\ }\href@noop {} {\bibfield  {journal}
  {\bibinfo  {journal} {Physical Review X}\ }\textbf {\bibinfo {volume} {11}},\
  \bibinfo {pages} {041036} (\bibinfo {year} {2021})}\BibitemShut {NoStop}%
\bibitem [{\citenamefont {Koczor}(2021)}]{koczor2021exponential}%
  \BibitemOpen
  \bibfield  {author} {\bibinfo {author} {\bibfnamefont {B.}~\bibnamefont
  {Koczor}},\ }\href@noop {} {\bibfield  {journal} {\bibinfo  {journal}
  {Physical Review X}\ }\textbf {\bibinfo {volume} {11}},\ \bibinfo {pages}
  {031057} (\bibinfo {year} {2021})}\BibitemShut {NoStop}%
\bibitem [{\citenamefont {Huo}\ and\ \citenamefont {Li}(2022)}]{huo2022dual}%
  \BibitemOpen
  \bibfield  {author} {\bibinfo {author} {\bibfnamefont {M.}~\bibnamefont
  {Huo}}\ and\ \bibinfo {author} {\bibfnamefont {Y.}~\bibnamefont {Li}},\
  }\href@noop {} {\bibfield  {journal} {\bibinfo  {journal} {Physical Review
  A}\ }\textbf {\bibinfo {volume} {105}},\ \bibinfo {pages} {022427} (\bibinfo
  {year} {2022})}\BibitemShut {NoStop}%
\bibitem [{\citenamefont {Yoshioka}\ \emph {et~al.}(2022)\citenamefont
  {Yoshioka}, \citenamefont {Hakoshima}, \citenamefont {Matsuzaki},
  \citenamefont {Tokunaga}, \citenamefont {Suzuki},\ and\ \citenamefont
  {Endo}}]{yoshioka2022generalized}%
  \BibitemOpen
  \bibfield  {author} {\bibinfo {author} {\bibfnamefont {N.}~\bibnamefont
  {Yoshioka}}, \bibinfo {author} {\bibfnamefont {H.}~\bibnamefont {Hakoshima}},
  \bibinfo {author} {\bibfnamefont {Y.}~\bibnamefont {Matsuzaki}}, \bibinfo
  {author} {\bibfnamefont {Y.}~\bibnamefont {Tokunaga}}, \bibinfo {author}
  {\bibfnamefont {Y.}~\bibnamefont {Suzuki}}, \ and\ \bibinfo {author}
  {\bibfnamefont {S.}~\bibnamefont {Endo}},\ }\href@noop {} {\bibfield
  {journal} {\bibinfo  {journal} {Physical Review Letters}\ }\textbf {\bibinfo
  {volume} {129}},\ \bibinfo {pages} {020502} (\bibinfo {year}
  {2022})}\BibitemShut {NoStop}%
\bibitem [{\citenamefont {Seif}\ \emph {et~al.}(2023)\citenamefont {Seif},
  \citenamefont {Cian}, \citenamefont {Zhou}, \citenamefont {Chen},\ and\
  \citenamefont {Jiang}}]{seif2023shadow}%
  \BibitemOpen
  \bibfield  {author} {\bibinfo {author} {\bibfnamefont {A.}~\bibnamefont
  {Seif}}, \bibinfo {author} {\bibfnamefont {Z.-P.}\ \bibnamefont {Cian}},
  \bibinfo {author} {\bibfnamefont {S.}~\bibnamefont {Zhou}}, \bibinfo {author}
  {\bibfnamefont {S.}~\bibnamefont {Chen}}, \ and\ \bibinfo {author}
  {\bibfnamefont {L.}~\bibnamefont {Jiang}},\ }\href@noop {} {\bibfield
  {journal} {\bibinfo  {journal} {PRX Quantum}\ }\textbf {\bibinfo {volume}
  {4}},\ \bibinfo {pages} {010303} (\bibinfo {year} {2023})}\BibitemShut
  {NoStop}%
\bibitem [{\citenamefont {O’Brien}\ \emph {et~al.}(2021)\citenamefont
  {O’Brien}, \citenamefont {Polla}, \citenamefont {Rubin}, \citenamefont
  {Huggins}, \citenamefont {McArdle}, \citenamefont {Boixo}, \citenamefont
  {McClean},\ and\ \citenamefont {Babbush}}]{o2021error}%
  \BibitemOpen
  \bibfield  {author} {\bibinfo {author} {\bibfnamefont {T.~E.}\ \bibnamefont
  {O’Brien}}, \bibinfo {author} {\bibfnamefont {S.}~\bibnamefont {Polla}},
  \bibinfo {author} {\bibfnamefont {N.~C.}\ \bibnamefont {Rubin}}, \bibinfo
  {author} {\bibfnamefont {W.~J.}\ \bibnamefont {Huggins}}, \bibinfo {author}
  {\bibfnamefont {S.}~\bibnamefont {McArdle}}, \bibinfo {author} {\bibfnamefont
  {S.}~\bibnamefont {Boixo}}, \bibinfo {author} {\bibfnamefont {J.~R.}\
  \bibnamefont {McClean}}, \ and\ \bibinfo {author} {\bibfnamefont
  {R.}~\bibnamefont {Babbush}},\ }\href@noop {} {\bibfield  {journal} {\bibinfo
   {journal} {PRX Quantum}\ }\textbf {\bibinfo {volume} {2}},\ \bibinfo {pages}
  {020317} (\bibinfo {year} {2021})}\BibitemShut {NoStop}%
\bibitem [{\citenamefont {Xiong}\ \emph {et~al.}(2020)\citenamefont {Xiong},
  \citenamefont {Chandra}, \citenamefont {Ng},\ and\ \citenamefont
  {Hanzo}}]{xiong2020sampling}%
  \BibitemOpen
  \bibfield  {author} {\bibinfo {author} {\bibfnamefont {Y.}~\bibnamefont
  {Xiong}}, \bibinfo {author} {\bibfnamefont {D.}~\bibnamefont {Chandra}},
  \bibinfo {author} {\bibfnamefont {S.~X.}\ \bibnamefont {Ng}}, \ and\ \bibinfo
  {author} {\bibfnamefont {L.}~\bibnamefont {Hanzo}},\ }\href@noop {}
  {\bibfield  {journal} {\bibinfo  {journal} {IEEE Access}\ }\textbf {\bibinfo
  {volume} {8}},\ \bibinfo {pages} {228967} (\bibinfo {year}
  {2020})}\BibitemShut {NoStop}%
\bibitem [{\citenamefont {Piveteau}\ \emph {et~al.}(2021)\citenamefont
  {Piveteau}, \citenamefont {Sutter}, \citenamefont {Bravyi}, \citenamefont
  {Gambetta},\ and\ \citenamefont {Temme}}]{piveteau2021error}%
  \BibitemOpen
  \bibfield  {author} {\bibinfo {author} {\bibfnamefont {C.}~\bibnamefont
  {Piveteau}}, \bibinfo {author} {\bibfnamefont {D.}~\bibnamefont {Sutter}},
  \bibinfo {author} {\bibfnamefont {S.}~\bibnamefont {Bravyi}}, \bibinfo
  {author} {\bibfnamefont {J.~M.}\ \bibnamefont {Gambetta}}, \ and\ \bibinfo
  {author} {\bibfnamefont {K.}~\bibnamefont {Temme}},\ }\href@noop {}
  {\bibfield  {journal} {\bibinfo  {journal} {Physical Review Letters}\
  }\textbf {\bibinfo {volume} {127}},\ \bibinfo {pages} {200505} (\bibinfo
  {year} {2021})}\BibitemShut {NoStop}%
\bibitem [{\citenamefont {Lostaglio}\ and\ \citenamefont
  {Ciani}(2021)}]{lostaglio2021error}%
  \BibitemOpen
  \bibfield  {author} {\bibinfo {author} {\bibfnamefont {M.}~\bibnamefont
  {Lostaglio}}\ and\ \bibinfo {author} {\bibfnamefont {A.}~\bibnamefont
  {Ciani}},\ }\href@noop {} {\bibfield  {journal} {\bibinfo  {journal}
  {Physical Review Letters}\ }\textbf {\bibinfo {volume} {127}},\ \bibinfo
  {pages} {200506} (\bibinfo {year} {2021})}\BibitemShut {NoStop}%
\bibitem [{\citenamefont {Figgatt}\ \emph {et~al.}(2019)\citenamefont
  {Figgatt}, \citenamefont {Ostrander}, \citenamefont {Linke}, \citenamefont
  {Landsman}, \citenamefont {Zhu}, \citenamefont {Maslov},\ and\ \citenamefont
  {Monroe}}]{figgatt2019parallel}%
  \BibitemOpen
  \bibfield  {author} {\bibinfo {author} {\bibfnamefont {C.}~\bibnamefont
  {Figgatt}}, \bibinfo {author} {\bibfnamefont {A.}~\bibnamefont {Ostrander}},
  \bibinfo {author} {\bibfnamefont {N.~M.}\ \bibnamefont {Linke}}, \bibinfo
  {author} {\bibfnamefont {K.~A.}\ \bibnamefont {Landsman}}, \bibinfo {author}
  {\bibfnamefont {D.}~\bibnamefont {Zhu}}, \bibinfo {author} {\bibfnamefont
  {D.}~\bibnamefont {Maslov}}, \ and\ \bibinfo {author} {\bibfnamefont
  {C.}~\bibnamefont {Monroe}},\ }\href@noop {} {\bibfield  {journal} {\bibinfo
  {journal} {Nature}\ }\textbf {\bibinfo {volume} {572}},\ \bibinfo {pages}
  {368} (\bibinfo {year} {2019})}\BibitemShut {NoStop}%
\bibitem [{\citenamefont {Takagi}(2021)}]{takagi2021optimal}%
  \BibitemOpen
  \bibfield  {author} {\bibinfo {author} {\bibfnamefont {R.}~\bibnamefont
  {Takagi}},\ }\href@noop {} {\bibfield  {journal} {\bibinfo  {journal}
  {Physical Review Research}\ }\textbf {\bibinfo {volume} {3}},\ \bibinfo
  {pages} {033178} (\bibinfo {year} {2021})}\BibitemShut {NoStop}%
\bibitem [{\citenamefont {Takagi}\ \emph
  {et~al.}(2022{\natexlab{a}})\citenamefont {Takagi}, \citenamefont {Endo},
  \citenamefont {Minagawa},\ and\ \citenamefont {Gu}}]{takagi2022fundamental}%
  \BibitemOpen
  \bibfield  {author} {\bibinfo {author} {\bibfnamefont {R.}~\bibnamefont
  {Takagi}}, \bibinfo {author} {\bibfnamefont {S.}~\bibnamefont {Endo}},
  \bibinfo {author} {\bibfnamefont {S.}~\bibnamefont {Minagawa}}, \ and\
  \bibinfo {author} {\bibfnamefont {M.}~\bibnamefont {Gu}},\ }\href@noop {}
  {\bibfield  {journal} {\bibinfo  {journal} {npj Quantum Information}\
  }\textbf {\bibinfo {volume} {8}},\ \bibinfo {pages} {114} (\bibinfo {year}
  {2022}{\natexlab{a}})}\BibitemShut {NoStop}%
\bibitem [{\citenamefont {Takagi}\ \emph
  {et~al.}(2022{\natexlab{b}})\citenamefont {Takagi}, \citenamefont {Tajima},\
  and\ \citenamefont {Gu}}]{takagi2022universal}%
  \BibitemOpen
  \bibfield  {author} {\bibinfo {author} {\bibfnamefont {R.}~\bibnamefont
  {Takagi}}, \bibinfo {author} {\bibfnamefont {H.}~\bibnamefont {Tajima}}, \
  and\ \bibinfo {author} {\bibfnamefont {M.}~\bibnamefont {Gu}},\ }\href@noop
  {} {\bibfield  {journal} {\bibinfo  {journal} {arXiv preprint
  arXiv:2208.09178}\ } (\bibinfo {year} {2022}{\natexlab{b}})}\BibitemShut
  {NoStop}%
\bibitem [{\citenamefont {Tsubouchi}\ \emph {et~al.}(2022)\citenamefont
  {Tsubouchi}, \citenamefont {Sagawa},\ and\ \citenamefont
  {Yoshioka}}]{tsubouchi2022universal}%
  \BibitemOpen
  \bibfield  {author} {\bibinfo {author} {\bibfnamefont {K.}~\bibnamefont
  {Tsubouchi}}, \bibinfo {author} {\bibfnamefont {T.}~\bibnamefont {Sagawa}}, \
  and\ \bibinfo {author} {\bibfnamefont {N.}~\bibnamefont {Yoshioka}},\
  }\href@noop {} {\bibfield  {journal} {\bibinfo  {journal} {arXiv preprint
  arXiv:2208.09385}\ } (\bibinfo {year} {2022})}\BibitemShut {NoStop}%
\bibitem [{\citenamefont {Hakoshima}\ \emph {et~al.}(2021)\citenamefont
  {Hakoshima}, \citenamefont {Matsuzaki},\ and\ \citenamefont
  {Endo}}]{hakoshima2021relationship}%
  \BibitemOpen
  \bibfield  {author} {\bibinfo {author} {\bibfnamefont {H.}~\bibnamefont
  {Hakoshima}}, \bibinfo {author} {\bibfnamefont {Y.}~\bibnamefont
  {Matsuzaki}}, \ and\ \bibinfo {author} {\bibfnamefont {S.}~\bibnamefont
  {Endo}},\ }\href@noop {} {\bibfield  {journal} {\bibinfo  {journal} {Physical
  Review A}\ }\textbf {\bibinfo {volume} {103}},\ \bibinfo {pages} {012611}
  (\bibinfo {year} {2021})}\BibitemShut {NoStop}%
\bibitem [{\citenamefont {Quek}\ \emph {et~al.}(2022)\citenamefont {Quek},
  \citenamefont {Fran{\c{c}}a}, \citenamefont {Khatri}, \citenamefont {Meyer},\
  and\ \citenamefont {Eisert}}]{quek2022exponentially}%
  \BibitemOpen
  \bibfield  {author} {\bibinfo {author} {\bibfnamefont {Y.}~\bibnamefont
  {Quek}}, \bibinfo {author} {\bibfnamefont {D.~S.}\ \bibnamefont
  {Fran{\c{c}}a}}, \bibinfo {author} {\bibfnamefont {S.}~\bibnamefont
  {Khatri}}, \bibinfo {author} {\bibfnamefont {J.~J.}\ \bibnamefont {Meyer}}, \
  and\ \bibinfo {author} {\bibfnamefont {J.}~\bibnamefont {Eisert}},\
  }\href@noop {} {\bibfield  {journal} {\bibinfo  {journal} {arXiv preprint
  arXiv:2210.11505}\ } (\bibinfo {year} {2022})}\BibitemShut {NoStop}%
\bibitem [{\citenamefont {Bomb{\'\i}n}(2015)}]{bombin2015single}%
  \BibitemOpen
  \bibfield  {author} {\bibinfo {author} {\bibfnamefont {H.}~\bibnamefont
  {Bomb{\'\i}n}},\ }\href@noop {} {\bibfield  {journal} {\bibinfo  {journal}
  {Physical Review X}\ }\textbf {\bibinfo {volume} {5}},\ \bibinfo {pages}
  {031043} (\bibinfo {year} {2015})}\BibitemShut {NoStop}%
\end{thebibliography}%

\appendix

\section{Robustness against the noise in ancilla qubits}
\label{sec_A1}
\begin{figure}[t]
    \begin{center}
        \includegraphics[width=0.9\linewidth]{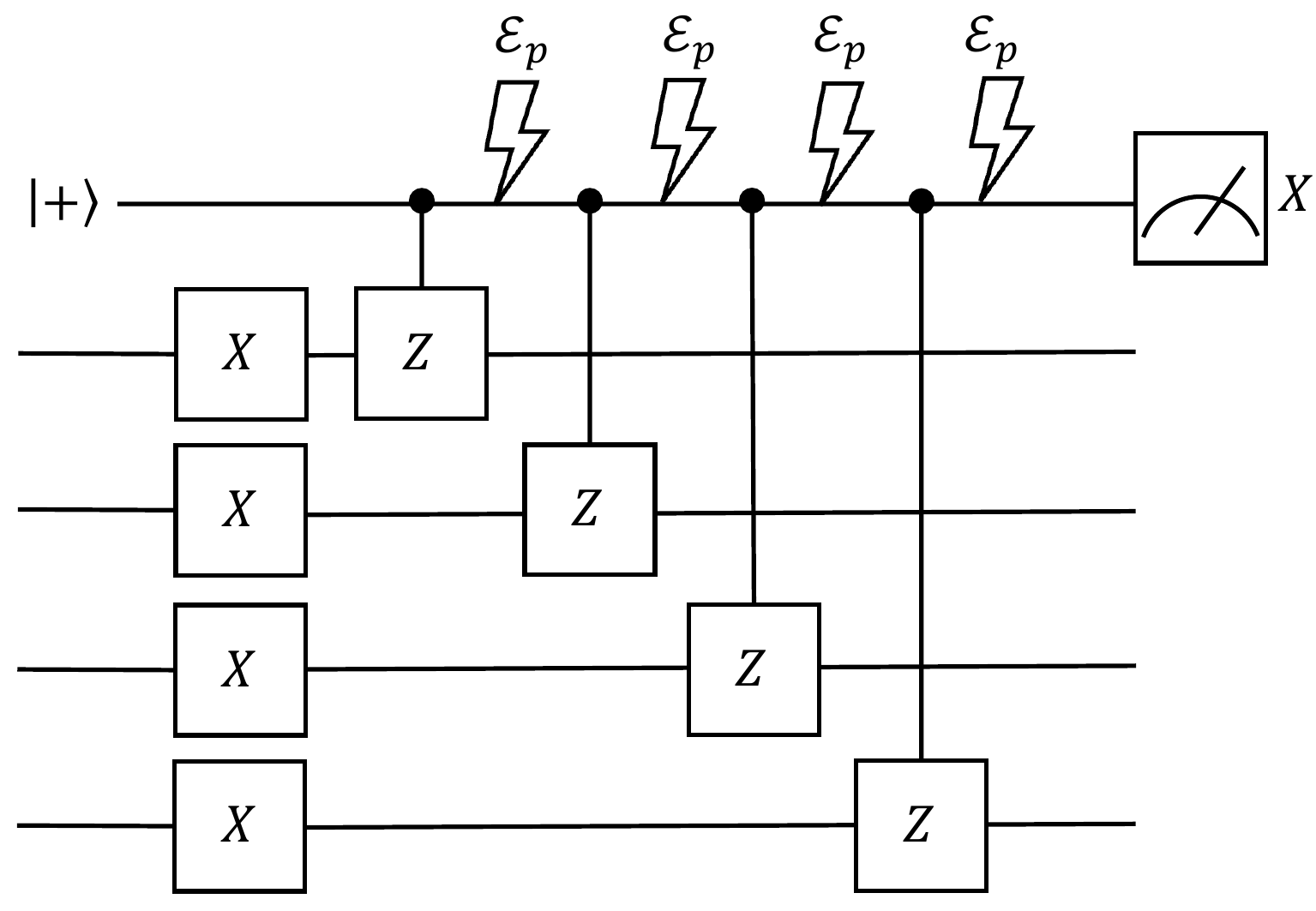}
        \caption{Gate-based decomposition of the VQED gadget in Fig. \ref{fig_vp_circuit} (b) for $[[4,1,2]]$ stabilizer code with $S_i = XXXX$ and $S_j = ZZZZ$.
        We assume that the single-qubit depolarizing noise $\mathcal{E}_p: \rho\mapsto (1-p)\rho + pI/2$ affects the ancilla qubit every time we perform controlled-Pauli gates.
        }
        \label{fig_vp_circuit_noisy}
    \end{center}
\end{figure}

In this section, we discuss the robustness of the VQED gadget against the noise in the ancilla qubits.
In order to perform $S_i$ or controlled-$S_j$ gates in the VQED gadget, we need to perform Pauli or controlled-Pauli gates as in Fig. \ref{fig_vp_circuit_noisy}.
Here, let us assume that the ancilla qubit is noisy and affected by the single-qubit depolarizing noise $\mathcal{E}_p: \rho\mapsto (1-p)\rho + pI/2$ every time we perform controlled-Pauli gates.
If the error occurred on the controlled-Pauli gates, the error propagates to the system qubits and may cause an undetectable error.
This may seem to ruin the performance of VQED.
However, our VQED protocol is not based on single-shot stabilizer measurements that are highly sensitive to such noise; our method only measures the expectation value of observables, and thus the effect of the error on the ancilla qubits can be removed.

When we apply the noisy VQED gadget as in Fig. \ref{fig_vp_circuit_noisy} to the state $\rho$, the state before the measurement will be
\begin{eqnarray}
\begin{aligned}
    \frac{(1-p)^{w(S_j)}}{2}&\big(\ketbra{0}\otimes S_i\rho S_i + \ketbra{1}\otimes S_jS_i\rho S_iS_j\\
    &+\ketbra{1}{0}\otimes  S_jS_i\rho S_i + \ketbra{0}{1}\otimes S_i\rho S_iS_j\big )\\
    &+\ketbra{0}\otimes\rho' + \ketbra{1}\otimes\rho'',
\end{aligned}
\end{eqnarray}
where ${w(S_j)}$ is the Pauli weight of $S_j$ (the number of Pauli operator in $S_j$) and $\rho'$ and $\rho''$ are unnormalized noisy quantum states in the system qubits.
By inserting additional noise in the ancilla qubits before the measurement, we can further convert this state to
\begin{eqnarray}
\label{eq_ap1}
    \begin{aligned}
    \frac{(1-p)^n}{2}&\big(\ketbra{0}\otimes S_i\rho S_i + \ketbra{1}\otimes S_jS_i\rho S_iS_j\\
    &+\ketbra{1}{0}\otimes  S_jS_i\rho S_i + \ketbra{0}{1}\otimes S_i\rho S_iS_j\big )\\
    &+\ketbra{0}\otimes\rho''' + \ketbra{1}\otimes\rho'''',
\end{aligned}
\end{eqnarray}
where $\rho'''$ and $\rho''''$ are also unnormalized noisy quantum states in the system qubits.
When we take the expectation value of the operator $X\otimes O$ for this state, we obtain
\begin{eqnarray}
    \frac{(1-p)^n}{2}(\tr[S_jS_i\rho S_iO] + \tr[S_i\rho S_iS_jO]),
\end{eqnarray}
and by uniformly sampling $i, j\in \qty{1,\cdots, 2^{n-k}}$ and taking the average of the distribution, we obtain
\begin{eqnarray}
    (1-p)^n \tr[P\rho PO].
\end{eqnarray}
Thus, even under the existence of noise on the ancilla qubit, we can still calculate the expectation value corresponding to the post-selected state $\rho_{\mathrm{det}}$ as
\begin{eqnarray}
    \tr[\rho_{\mathrm{det}}O]= \frac{(1-p)^n\tr[P\rho P O]}{(1-p)^n\tr[P\rho P]}.
\end{eqnarray}

In the same way, we can say that the value obtained through VQED in Eq. (\ref{eq_vqed}) remains unchanged even under the existence of such noise, since the numerator and the denominator of Eq. (\ref{eq_vqed}) are only multiplied by $(1-p)^{nL}$.
The only change to the performance of VQED is the slight increase in the sampling overhead from $N = O(\varepsilon^{-2} \tr[\rho_{\mathrm{det}}']^{-2})$ to $N = O(\varepsilon^{-2} (1-p)^{-2nL}\tr[\rho_{\mathrm{det}}']^{-2})$.

The essential point of this robustness is that the noisy term $\ketbra{0}\otimes\rho''' + \ketbra{1}\otimes\rho''''$ in Eq. (\ref{eq_ap1}) is removed when we take the expectation value of $X$ for the ancilla qubit.
We note that the same principle applies to other noise models that are not biased by Pauli $X$ or $Y$, such as local dephasing and amplitude damping noise.
We also note that the noise model we used for the VQED gadget in our numerical simulation is different from what we consider in this section: instead of assuming that the noise affects the ancilla qubits every after the execution of the controlled-Pauli gates, we assumed that the noise affects both the system and ancilla qubits every after the execution of $S_i$ and controlled-$S_j$ gates.

\begin{figure*}[t]
    \begin{center}
        \resizebox{0.7\hsize}{!}{\includegraphics{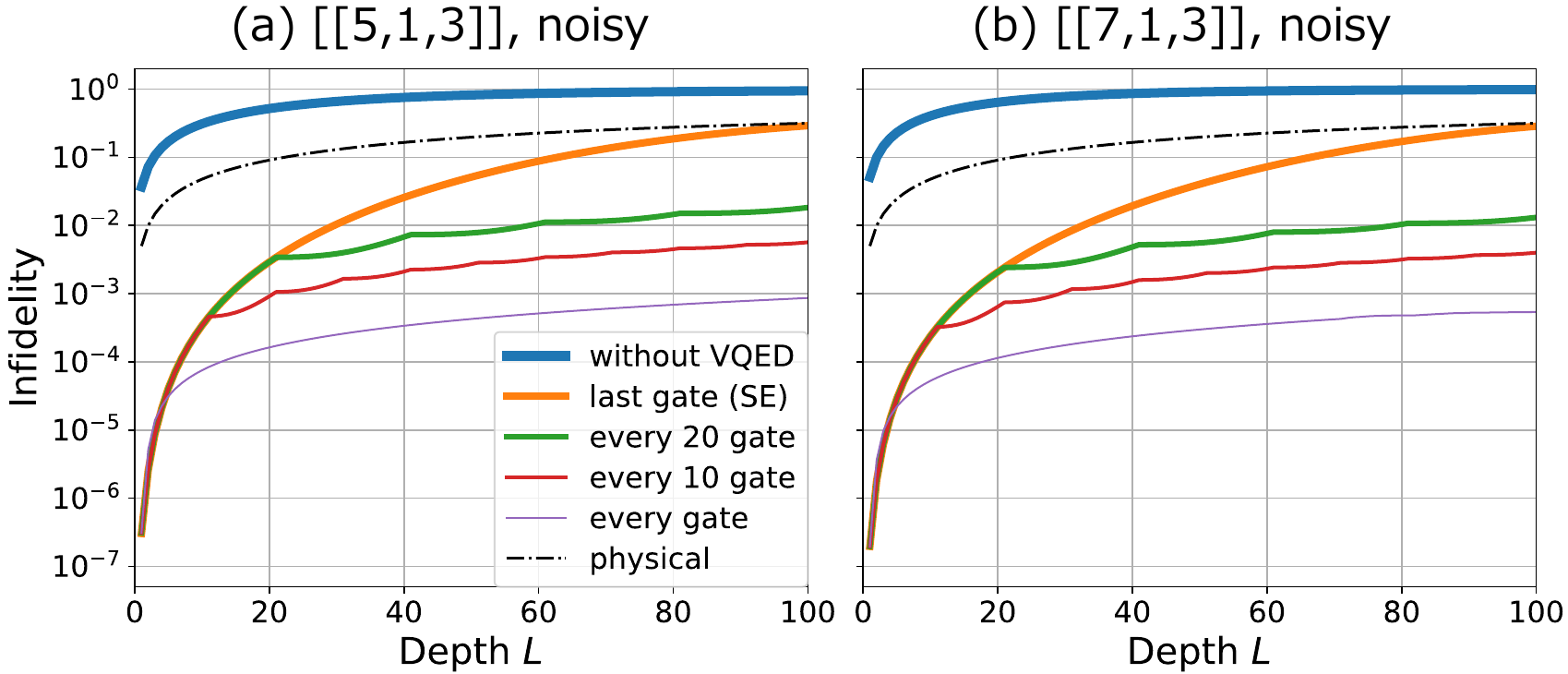}}
        \caption{Depth $L$ dependence of infidelity $1-\mathrm{tr}[\rho_{\mathrm{det}}\ketbra{\bar{\Psi}}] = 1-\ev*{\rho_{\mathrm{det}}}{\bar{\Psi}}$ between the output state of the noisy circuit with VQED $\rho_{\mathrm{det}}$ and the noiseless circuit $\ket*{\bar{\Psi}}$ for (a): \red{$[[5,1,3]]$} and (b): \red{$[[7,1,3]]$} stabilizer codes. 
        All of the panels denote the results when the VQED gadgets are affected by local depolarizing noise.
        The ``without \red{VQED}'' line represents the infidelity when we did not perform VQED. The ``last gate \red{(SE)}'' line represents infidelity when we perform VQED only before the measurement\red{, which is just a normal SE,} as in Refs. \cite{mcclean2020decoding,cai2021quantum}. The ``every 20 gates'' and the ``every 10 gates'' lines represent infidelity when we perform VQED after every 20 and 10 gates. The ``every gate'' line represents infidelity when we perform VQED after every gate. The ``physical'' line represents the infidelity of a single physical qubit without encoding.
        }
        \label{fig_infidelity_ap}
    \end{center}
\end{figure*}

\begin{figure*}[t]
    \begin{center}
        \resizebox{0.7\hsize}{!}{\includegraphics{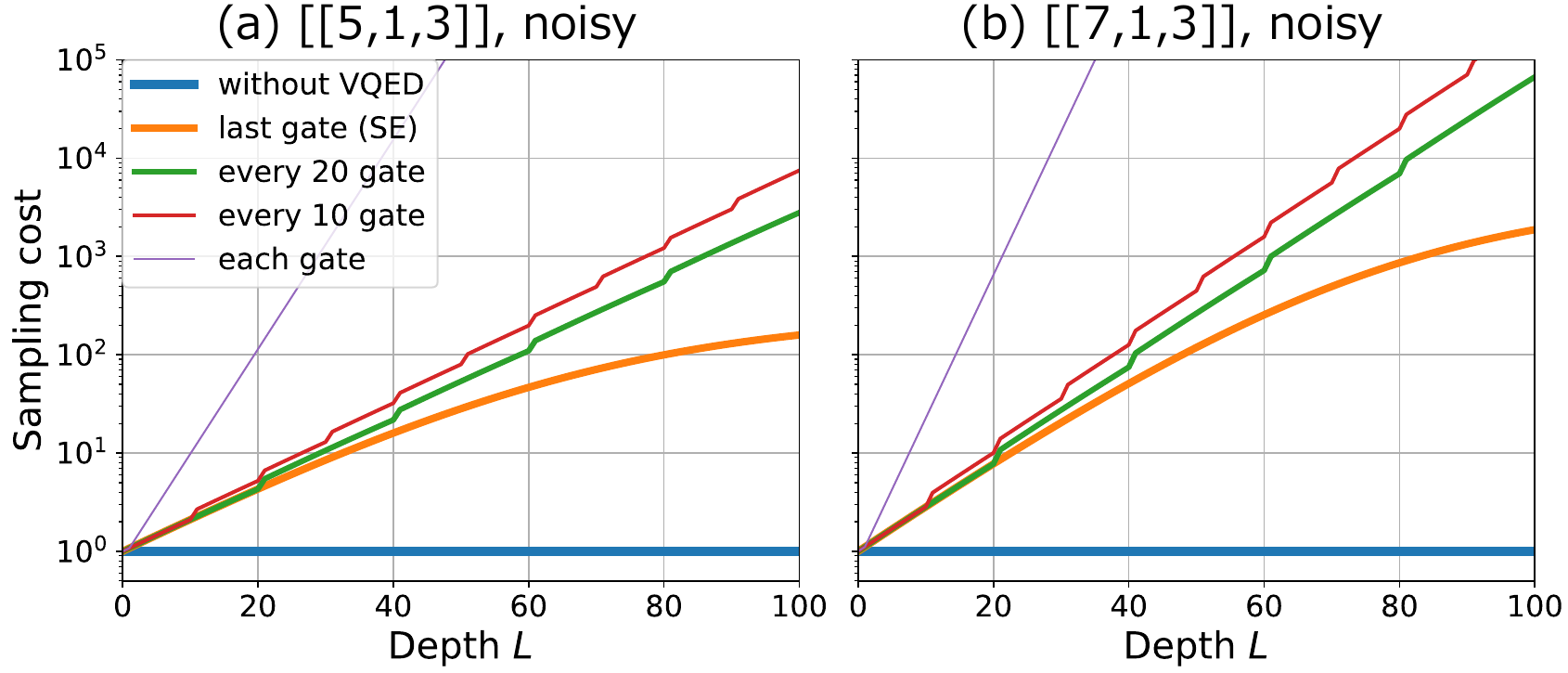}}
        \caption{Scaling of the sampling cost $\mathrm{\tr}[\rho'_{\mathrm{det}}]^{-2}$ with respect to the depth $L$ of the quantum circuit for (a): \red{$[[5,1,3]]$} and (b): \red{$[[7,1,3]]$} stabilizer codes. 
        All of the panels denote the results when the VQED gadgets are affected by local depolarizing noise.
        The ``without \red{VQED}'' line represents \red{sampling cost} when we did not perform VQED. The ``last gate \red{(SE)}'' line represents \red{sampling cost} when we perform VQED only before the measurement\red{, which is just a normal SE,} as in Refs. \cite{mcclean2020decoding,cai2021quantum}. The ``every 20 gates'' and the ``every 10 gates'' lines represent the sampling cost when we perform VQED after every 20 and 10 gates. The ``every gate'' line represents the sampling cost when we perform VQED after every gate.
        }
        \label{fig_cost_ap}
    \end{center}
\end{figure*}

\section{Transversal single-qubit gates in stabilizer codes}
\label{sec_A2}
In this section, we clarify the sets of transversal single-qubit gates we use in the numerical simulation.
For $[[4, 1, 2]]$ stabilizer code, we use a set of single-qubit Pauli gates as a set of transversal single-qubit gates~\cite{gottesman2016quantum}.
For $[[5, 1, 3]]$ stabilizer code, we use $\qty{X, Y, Z, SH}$ as a set of transversal single-qubit gates~\cite{gottesman1997stabilizer}.
For $[[7, 1, 3]]$ stabilizer code, we use a set of single-qubit Clifford gates as a set of transversal single-qubit gates~\cite{gottesman1997stabilizer}.

\section{Numerical simulation of VQED when the VQED gadgets are noisy}
\label{sec_A3}
In this section, we present the performance of VQED for $[[5, 1, 3]]$ and $[[7, 1, 3]]$ stabilizer codes when the VQED gadget $S_{i_l}$ gate and controlled-$S_{j_l}$ gate in Fig. \ref{fig_VQED_circuit} are each affected by the local depolarizing noise $\mathcal{E}_{p}^{\otimes n}$ and $\mathcal{E}_{p}^{\otimes (n+1)}$ with the same error rate $p=0.01$.
Our results are shown in Fig. \ref{fig_infidelity_ap} and Fig. \ref{fig_cost_ap}.
The results for $[[5, 1, 3]]$ and $[[7, 1, 3]]$ stabilizer codes are qualitatively similar to those for the $[[4,1,2]]$ stabilizer codes shown in Fig. \ref{fig_infidelity} (d) and Fig. \ref{fig_cost} (d).

\newpage

\end{document}